\documentclass[letterpaper]{article} % DO NOT CHANGE THIS
\usepackage{aaai24}  % DO NOT CHANGE THIS
\usepackage{times}  % DO NOT CHANGE THIS
\usepackage{helvet}  % DO NOT CHANGE THIS
\usepackage{courier}  % DO NOT CHANGE THIS
\usepackage[hyphens]{url}  % DO NOT CHANGE THIS
\usepackage{graphicx} % DO NOT CHANGE THIS
\usepackage{natbib}  % DO NOT CHANGE THIS AND DO NOT ADD ANY OPTIONS TO IT
\usepackage{caption} % DO NOT CHANGE THIS AND DO NOT ADD ANY OPTIONS TO IT
\usepackage{booktabs}
\usepackage{array}
\usepackage{multirow}
\usepackage{color}
\definecolor{lightgray}{RGB}{215,215,215}
\usepackage{colortbl}  %彩色表格需要加载的宏包
\usepackage{xcolor}
\usepackage{subfigure}
\usepackage{algorithm}  
\usepackage{algorithmicx}  
\usepackage[noend]{algpseudocode}  

\usepackage{amsmath}  
\usepackage{amssymb}
\usepackage{bm}
\usepackage{enumitem}
\usepackage{tabularx}
\usepackage[utf8]{inputenc}
\usepackage[english]{babel}
\usepackage{amsthm}
\usepackage{bm}
\newcommand{\ie}{\emph{i.e., }}
\newcommand{\eg}{\emph{e.g., }}

\newcommand{\wrt}{\emph{w.r.t. }}
\newcommand{\cf}{\emph{cf. }}
\newcommand{\aka}{\emph{a.k.a. }}
\DeclareUnicodeCharacter{2217}{*}

\newlength\myindent
\floatname{algorithm}{Algorithm}

\usepackage{newfloat}
\usepackage{listings}
\DeclareCaptionStyle{ruled}{labelfont=normalfont,labelsep=colon,strut=off} % DO NOT CHANGE THIS
\floatstyle{ruled}
\newfloat{listing}{tb}{lst}{}
\floatname{listing}{Listing}
\title{Temporally and Distributionally Robust Optimization for Cold-Start Recommendation}
\author {
    Xinyu Lin\textsuperscript{\rm 1},
    Wenjie Wang\textsuperscript{\rm 1}\thanks{Corresponding author. This work is supported by the National Key Research and Development Program of China (2022YFB3104701), the National Natural Science Foundation of China (62272437), and Huawei International Pte Ltd.},
    Jujia Zhao\textsuperscript{\rm 1},
    Yongqi Li\textsuperscript{\rm 2},
    Fuli Feng\textsuperscript{\rm 3},
    Tat-Seng Chua\textsuperscript{\rm 1}
}
\affiliations {
    \textsuperscript{\rm 1}National University of Singapore\\
    \textsuperscript{\rm 2}The Hong Kong Polytechnic University\\
    \textsuperscript{\rm 3}MoE Key Laboratory of Brain-inspired Intelligent Perception and Cognition, University of Science and Technology of China\\
    {\{xylin1028, wenjiewang96, zhao.jujia.0913, liyongqi0, fulifeng93\}}@gmail.com, dcscts@nus.edu.sg
}

\begin{document}

\maketitle

\begin{abstract}
% Traditional Collaborative Filtering (CF) recommender models highly depend on user-item interactions to learn CF representations. 
% Such CF methods often fall short of recommending cold-start items due to the lack of historical interactions. 
Collaborative Filtering (CF) recommender models highly depend on user-item interactions to learn CF representations, thus falling short of recommending cold-start items. 
To address this issue, prior studies mainly introduce item features (\eg thumbnails) for cold-start item recommendation. 
They learn a feature extractor on warm-start items to align 
% between the feature representations and interactions, 
feature representations with interactions, 
and then leverage the feature extractor to 
extract the feature representations of cold-start items for interaction prediction.  
Unfortunately, 
the features of cold-start items, especially the popular ones, tend to diverge from those of warm-start ones due to temporal feature shifts, 
preventing the feature extractor from accurately learning feature representations of cold-start items. 

% To alleviate the impact of temporal item feature shifts, we employ Distributionally Robust Optimization (DRO), which has the potential to confer robustness whereas encounters the inconsistency issue for the cold-start recommendation. 
To alleviate the impact of temporal feature shifts, we consider using Distributionally Robust Optimization (DRO) to enhance the generation ability of the feature extractor. Nonetheless, existing DRO methods face an inconsistency issue: 
% the minority groups emphasized during DRO training might not align with the cold-start item distribution. 
the worse-case warm-start items emphasized during DRO training might not align well with the cold-start item distribution. 
To capture the temporal feature shifts and combat this inconsistency issue, 
we propose a novel temporal DRO with new optimization objectives, namely, 1) to 
integrate a worst-case factor to improve the worst-case performance, 
% optimization, thereby raising the lower bound of performance, 
and 2) to devise a shifting factor to capture the shifting trend of item features and enhance the optimization of the potentially popular groups in cold-start items. 
% We instantiate temporal DRO on two competitive cold-start recommender models and conduct substantial experiments on three real-world datasets under various evaluation settings. 
% (\eg cold- and warm-start recommendation, and recommendation under differing degrees of feature shifts). 
% Empirical results 
Substantial experiments on three real-world datasets
validate the superiority of our temporal DRO in enhancing the generalization ability of cold-start recommender models. 
% \vspace{-0.1cm}
\end{abstract}

\section{Introduction}
\label{sec:introduction}

% p1
% Recommender systems aim to help users filter the overloaded multimedia information on the web to meet their personalized information needs.  
% Generally speaking, the most representative methods of traditional recommender systems utilize Collaborative Filtering (CF) for personalized recommendation. In essence, CF methods learn CF representations of user and item from historical interactions and utilize CF representations to predict the user preference over all items. 
% Meanwhile, numerous new items such as micro-videos are continuously being created and iterated for recommender systems to generate personalized recommendation. For example, 500 hours of video are uploaded to YouTube every minute and more than one million micro-videos are uploaded to TikTok everyday. 
% However, since the newly-uploaded items (a.k.a, cold-start items)  lack historical interactions with any users and thus have no CF representations, traditional recommender systems usually fail to recommend cold-start items appropriately for users. 
% Hence, it is of vital importance to devise efficient algorithms for cold-start item prediction for recommender systems.

Recommender systems are widely deployed to filter the overloaded multimedia information on the web for meeting users' personalized information needs~\cite{He2017Neural}. 
% Generally speaking, the most representative method of recommender models utilizes Collaborative Filtering (CF) for personalized recommendation~\cite{koren2009matrix}. 
% Generally speaking, Collaborative Filtering (CF) is the most commonly used method of recommender models for personalized recommendation~\cite{koren2009matrix}. 
Technically speaking, Collaborative Filtering (CF) is the most representative method~\cite{koren2009matrix}. 
In essence, CF methods learn the CF representations of users and items from historical interactions and utilize the learned CF representations to predict the users' future interactions. 
% preference over items. 
% With the continual advancement of production capabilities, recommender systems are challenged by the ever-increasing influx of new items (\aka cold-start items\footnote{For simplicity, the cold-start items and warm-start items are shorted as cold and warm items.}). 
As content production capabilities continue to advance, recommender systems face the challenge of accommodating an increasing influx of new items (\aka cold-start items\footnote{For simplicity, cold-start items and warm-start items are referred to as cold and warm items, respectively.}). 
For example, 
% an estimated 
500 hours of video are uploaded to YouTube every minute\footnote{https://www.statista.com/.}. 
% and more than one million micro-videos are uploaded to TikTok every day. 
% However, s
Since the new items lack historical interactions and thereby have no CF representations, 
% traditional CF recommender systems usually fail to effectively recommend these cold items for users, disrupting the ecological balance of recommender systems on the item side.
traditional CF methods fail to effectively recommend these cold items to 
% for 
users, disrupting the ecological balance of recommender systems on the item side. 
In light of this, 
% it is crucial to 
% devise 
% efficient algorithms for 
it is essential to improve the cold-start item recommendation. 
Prior literature has integrated item features, such as categories and thumbnails of micro-videos, for cold-start item recommendation~\cite{shalaby2022m2trec,zhao2022improving}. 
% Typically, existing methods essentially learn a feature extractor that encodes items into feature representations and utilizes feature representations to fit the user-item interactions. 
These methods essentially learn a feature extractor that encodes warm items (\ie items in the training set) into feature representations and utilizes feature representations to fit the user-item interactions during training. 
% During inference for cold items, only the feature representations of the item is leveraged for user preference estimation in the absence of CF counterparts. 
% For inference for cold items, only their feature representations are utilized to estimate user preference, given the lack of CF counterparts. 
% For recommending cold items, only feature representations from the feature extractor are utilized to estimate user preference, given the lack of CF counterparts. 
For inference for cold items, given the lack of CF counterparts, only feature representations from the feature extractor are used to estimate user preference. 
% Following this paradigm, previous works mainly devise different strategies to align feature representations and user-item interactions, which mainly fall into two lines. 1) Robust training-based methods~\cite{volkovs2017dropoutnet,du2020learn} use both feature representations and CF representations to predict interactions while CF representations is randomly corrupted to strengthen the alignment between feature representations and interactions. 
The key of this paradigm lies in devising training strategies to align feature representations and user-item interactions, which mainly fall into two research lines. 
% 1) Robust training-based methods~\cite{volkovs2017dropoutnet,du2020learn} use both feature representations and CF representations to predict interactions while CF representations are randomly corrupted to strengthen the alignment between feature representations and interactions. 
1) Robust training-based methods~\cite{volkovs2017dropoutnet,du2020learn} use both feature representations and CF representations to predict interactions while CF representations are randomly corrupted to strengthen the alignment. 
2) Auxiliary loss-based methods~\cite{zhu2020recommendation} pay attention to minimizing the distance between the feature representations and CF representations learned from interactions via the auxiliary loss, \eg contrastive loss~\cite{wei2021contrastive} and GAN loss~\cite{chen2022generative}.
% , and Wasserstein Distance~\cite{zhao2022improving}.

% For example, CLCRec~\cite{} maps the item feature into the feature representations and pushes it to be closer to the CF space via contrastive learning. GAR~\cite{} utilizes adversarial training to learn a robust feature extractor, bridging the gap between CF space and feature representations space. Moreover, a recently proposed CCFCRec~\cite{} leverages co-occurrence of warm-start items for contrastive learning to mitigate the blurry issue of item feature representations, which is caused by inappropriate alignment between warm-start item and cold-start item.

\begin{figure}[t]
% \vspace{-0.3cm}
% \setlength{\abovecaptionskip}{0cm}
% \setlength{\belowcaptionskip}{0cm}
  % \centering  
  \begin{center}

  \hspace{-0.105in} 
  % \subfigure{
  %   \includegraphics[height=1.6in]{figures/intro_shift_a_tp.pdf}}
  % \hspace{-0.105in}
  % \subfigure{
  %    \includegraphics[height=1.6in]{figures/intro_shift_b_tp.pdf}}
\subfigure{
    \includegraphics[height=2.3in]{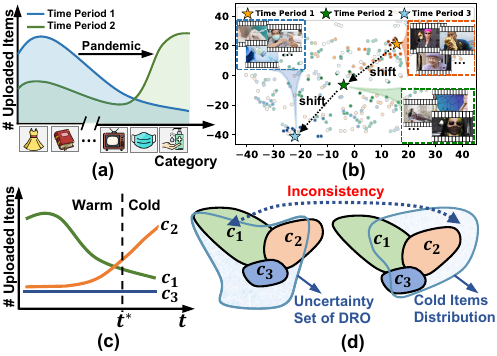}
}
  % \hspace{-0.105in}
%   \vspace{-0.1in}
% \vspace{-0.1cm}
\caption{(a) An example of item category feature shifts towards sanitary products. 
(b) T-SNE visualization of visual features of item thumbnails in three time periods on a Micro-video dataset. The stars represent the average item features in each time period. (c) An example of the shifting trend of three item groups over time. (d) Illustration of the inconsistency issue of DRO.}
% , and we randomly select one hundred items in each time period for visualization.
  \label{fig:intro_shift}
  % \vspace{-0.10cm}
\end{center}
\end{figure}

% p3
% However, existing methods inevitably suffer from temporal item feature shifts from warm items to cold items. Figure~\ref{fig:intro_shift}(a) illustrates the temporal shifts of item features due to some environmental factors, \eg the breakout of pandemic. Empirical evidence on the Micro-video dataset is also given in Figure~\ref{fig:intro_shift}(b), where item features are gradually shifting over time. 
% Since the feature extractor is trained on warm items via Empirical Risk Minimization (ERM), it can easily overfit the majority group of warm items. Whereas, the majority group of cold items can be different from that of warm items as shown in Figure~\ref{fig:intro_shift}, thus leading to poor cold item performance. 
% In this context, we require the cold-start recommender model to learn a feature extractor with a strong generalization ability that can achieve good performance on cold item distributions temporally shifted from warm items. 

% However, existing methods inevitably suffer from temporal item feature shifts from warm items to cold items. 
Despite their success, existing methods suffer from a critical issue: 
% temporal feature shifts from warm items to cold items. 
item features temporally shift from warm to cold items~\cite{wang2023equivariant}. 
% As shown in Figure~\ref{fig:intro_shift}(a), item feature shifts over time due to various environmental factors, such as the outbreak of a pandemic. 
As illustrated in Figure~\ref{fig:intro_shift}(a), the category features of newly-uploaded items are shifting over time due to various environmental factors, such as a pandemic outbreak. 
% Empirical evidence on the Micro-video dataset further confirms this phenomenon, as depicted in Figure~\ref{fig:intro_shift}(b): we first divide the items into three environments according to the uploading time, and then visualize the item features (visual features of thumbnails), where the star represents the average item features in each environment. It is observed that item features are gradually shifting over time, \ie the orange star shifts to the green, and then the blue star. 
Empirical evidence from a real-world Micro-video dataset further substantiates this phenomenon. In Figure~\ref{fig:intro_shift}(b), we divide the micro-videos into three time periods according to the upload time and visualize the micro-video features, where a star represents the average item features in each time period. The moving stars across time periods validate that item features are gradually shifting over time. 
% Since the feature extractor is trained on warm items via Empirical Risk Minimization (ERM)~\cite{vapnik1991principles}, 
% it may easily overfit the majority group of warm items. Unfortunately, the majority group of cold items can be different from that of warm items as shown in Figure~\ref{fig:intro_shift}, hindering the feature extractor from accurately extracting feature representations for interaction prediction. 
Since the feature extractor is typically trained on warm items using Empirical Risk Minimization (ERM)~\cite{vapnik1991principles}, 
it easily overfits the majority group of warm items. Unfortunately, the majority group of cold items could deviate from that of warm items as depicted in Figure~\ref{fig:intro_shift}(a) and (b). Such temporal feature shifts hinder the feature extractor's ability to accurately extract feature representations for cold items, thus degrading the performance of cold-start item recommendation. 
% In this context, the key to enhancing cold-start item recommendation is to learn a feature extractor with a strong generalization ability that can achieve good performance on temporally shifted cold item distributions. 
% To tackle the issue of item feature shifts, the key is to learn a feature extractor with robust generalization abilities, enhancing interaction prediction on temporally shifted cold items. 
To tackle this issue, we consider learning a feature extractor with robust generalization ability to enhance the interaction prediction on temporally shifted cold items. 

To strengthen the generalization ability, 
% To this end, 
Distributionally Robust Optimization (DRO) is a promising approach\footnote{Other potential solutions are discussed in Section~\ref{sec:experiment}.}. 
In general, DRO aims to
% achieve uniformly good performance by optimizing 
enhance 
the worst-case performance over the pre-defined uncertainty set, \ie potential shifted distributions~\cite{duchi2018learning}. 
% , where the uncertainty set contains mixtures of subgroups (\eg item groups) of training distribution. By optimizing the worst-case performance over the pre-defined uncertainty set, DRO can guarantee the tail performance over potential shifted distributions. 
However, directly applying DRO
% \footnote{We adopt the setting of Group DRO, which is a more practical approach of DRO family to avoid the over-pessimistic issue~\cite{oren2019distributionally}, where the training distribution is assumed to be a mixture of subgroups and the uncertainty set is defined on mixtures of these subgroups (\cf Section~\ref{sec:method}).} 
in cold-start recommendation suffers from the inconsistency issue. 
DRO will overemphasize the minority groups\footnote{Minority group usually yields worse performance in recommendation~\cite{wen2022distributionally}. In DRO, the training distribution is assumed to be a mixture of subgroups, and the uncertainty set is defined on mixtures of these subgroups (\cf Section~\ref{sec:preliminaries}).} in warm items at the expense of other groups' performance~\cite{oren2019distributionally}. 
% Due to the fact that the minority group over the warm items may not necessarily become popular in upcoming cold items, the overemphasis on the minority group of warm items might compromise the model expressiveness of some popular groups in cold items, limiting the robustness enhancement under temporal item feature shifts. 
Due to the fact that minority groups in warm items may not guarantee their popularity in subsequent cold items, the overemphasis on the minority group of warm items might compromise the performance of the popular groups in cold items. 
% , limiting the robustness enhancement of the feature extractor. 
For example, in Figure~\ref{fig:intro_shift}(c), $c_1$, $c_2$, and $c_3$ denote three item groups, where $c_3$ is the minority group in the warm items that traditional DRO 
% tries to improve the performance on. 
pays special attention to. 
% at the expense of the performance of $c_1$ and $c_2$. 
% Nevertheless, $c_2$ has gradually become popular over time and dominated the cold items. 
However, $c_2$ is gradually becoming popular, dominating the cold items. 
% As such, the inconsistency between the overemphasis over $c_3$ and the shifting trend towards $c_2$ leads to less satisfying mitigation of item feature shifts (see Figure~\ref{fig:shifting_trend_DRO_comparison}(b)). 
% Such inconsistency between the overemphasis over $c_3$ and the shifting trend towards $c_2$ leads to less satisfying mitigation of item feature shifts (see Figure~\ref{fig:shifting_trend_DRO_comparison}(b)). 
The inconsistency between the excessive emphasis on $c_3$ and the shifting trend towards $c_2$
% results in less satisfactory alleviation of item feature shifts (see Figure~\ref{fig:intro_shift}(d)). 
prevents DRO from alleviating the impact of temporal feature shifts (see Figure~\ref{fig:intro_shift}(d)). 
% 举例：（看图说话）
% To address this inconsistency issue and strengthen the generalization ability of the feature extractor against the temporal feature shift, we propose two objectives for DRO training: 
To address this inconsistency issue and strengthen the generalization ability of the feature extractor under the temporal feature shifts, we put forth two objectives for DRO training: 
% 1) enhance the worst-case performance of the minority group in warm items to guarantee the performance lower bound; meanwhile
1) enhancing the worst-case optimization on the minority group of warm items, thereby raising the lower bound of performance; and
2) capturing the shifting trend of item features and emphasizing the optimization of the groups likely to become popular.

To this end, we propose a \textbf{T}emporal \textbf{DRO} (TDRO), which 
% utilizes temporal shifting trend to enhance the performance beyond minority groups of warm items. 
considers the temporal shifting trend of item features for cold-start recommendation. 
% (用shifting trend来guide DRO的A)
% Specifically, 
% For data processing, 
% we divide the warm items into groups by
% % item features (\eg visual feature of thumbnails). 
% the similarity of item features (\eg visual feature of thumbnails). 
% todo: check the above sentence. 
% Besides, we sort the training interactions according to the timestamps and 
% % split them into different time periods, chronologically. 
% partition them into distinct time periods, chronologically. 
% During training, we consider two factors when applying DRO for model optimization: 
In particular, we consider two factors for the training of TDRO: 
% 1) \textit{Worst case factor} for worst-case performance guarantee, where we prioritize improving performance on item groups that have large training loss. 
1) \textit{a worst-case factor} to guarantee worst-case performance, where we divide the warm items into groups by the similarity of item features, 
% (\eg visual feature of thumbnails)
and prioritize the improvements of the item groups with large training loss; 
% 2) \textit{Shifting trend factor} for performance enhancement over popular groups in cold items. 
% To consider the temporal shifts of item features, we exploit time period-wise information to capture the shifting trend over temporally shifted distributions. 
% Besides, we devise an assisting strategy at each update step for TDRO: train on the groups that could lead to the best performance over the potential shifted distribution reflecting the shifting trend. 
and 2) \textit{a shifting factor} to capture the shifting trend of item features, which utilizes a gradient-based strategy to emphasize the optimization towards the gradually popular item groups across time periods. 
We instantiate the TDRO on two State-Of-The-Art (SOTA) cold-start recommender methods and conduct extensive experiments on three real-world datasets.  
The empirical results under multiple settings (\eg cold-start and warm-start recommendation, and recommendation with differing degrees of temporal feature shifts) validate the superiority of TDRO in enhancing the generalization ability of cold-start models. 
% while achieving 
% improvements on warm items. 
% superior performance on warm items. 
We release our codes and datasets at~\url{https://github.com/Linxyhaha/TDRO/}.

The contributions of this work are summarized as follows. 
\begin{itemize}[leftmargin=*]
    % \item We point out the temporal item feature shift issue in the cold-start item recommendation and propose to strengthen the generalization ability of the feature extractor.
    
    % \item We extend conventional DRO to avoid overemphasizing minority groups and leverage temporal shifting trend to yield more accurate predictions for groups that are likely to be popular in the cold items.
    
    % \item We conduct substantial experiments on three datasets, verifying the effectiveness of TDRO in achieving robust prediction against temporal item feature shift in cold-start recommendation.
    
    \item We emphasize the vulnerability of ERM and underscore the necessity of adjusting the learning objective to strengthen the generalization ability of cold-start models under temporal feature shifts. 
    \item We propose a novel TDRO objective for cold-start recommendation, which extends the conventional DRO to avoid overemphasizing the minority groups and capture the temporal shifting trend of item features. 
    \item We conduct extensive experiments on three datasets, demonstrating the effectiveness of temporal DRO in attaining robust prediction under temporal feature shifts. 
\end{itemize}
\section{Related Work}
\label{sec:related_work}

% \subsection{Cold-start Recommendation}
\noindent$\bullet\quad${\textbf{Cold-start Recommendation.}}
Traditional CF methods typically rely on CF representations learned from historical interactions~\cite{wang2022causal,li2019routing,sun2022counterfactual}. 
However, the influx of cold items hinders traditional CF methods from providing appropriate recommendations due to the lack of historical interactions~\cite{zhao2022improving,rajapakse2022fast,raziperchikolaei2021shared,pulis2021siamese,du2022socially,huan2022industrial,zhu2021fairness,sun2021form,wang2021privileged,chu2023meta}. 
% As for cold users, their interests can be effectively explored by recommending popular items to them. 
% Randomly recommending numerous cold items to learn item CF representations is inefficient and may hurt the user experience. 
% Therefore, it is crucial to solve the cold-start item issue and achieve accurate interaction prediction over cold items. 
% Hence, it is crucial to enhance the cold-start item recommendation~\cite{houlsby2014cold,pan2019warm,vartak2017meta,neupane2022dynamic,wei2020fast}. 
To remedy this, 
% Existing methods typically address the cold-start problem by aligning the extracted feature representations with CF counterparts in the following ways: (1) Alignment via robust training, where alignment is achieved by randomly dropping out the CF representation of warm items for robust prediction purely based on feature representation. For example, DropoutNet~\cite{volkovs2017dropoutnet} employs a technique of randomly dropping CF representations of warm items for more robust training, which can help to fill the gap and compensate for the missing CF information for cold items.  Instead of dropping CF representations, MTPR~\cite{du2020learn} uses a method of replacing warm item CF representations with zero vectors and applies multi-task learning for more robust training. (2) Alignment via auxiliary loss which explicitly constrain the distance between CF representation and feature representation by the auxiliary loss. For example, CLCRec~\cite{wei2021contrastive} maps the item feature into the feature representation and pushes it to be closer to the CF space via contrastive learning. GAR~\cite{chen2022generative} utilizes adversarial training to learn a robust feature extractor, bridging the gap between CF space and feature representation space.
existing methods align the feature representations with interactions~\cite{meng2020wasserstein,guo2017integration}, falling into two research lines. 
% 1) Robust training-based methods, where both feature and CF representations are utilized to predict interactions while the CF representations are randomly corrupted to encourage alignment~\cite{volkovs2017dropoutnet}. 
1) Robust training-based methods utilize both feature and CF representations for prediction while the CF representations are randomly corrupted~\cite{volkovs2017dropoutnet}. 
% between feature representations and interactions
% For example, DropoutNet~\cite{volkovs2017dropoutnet} randomly drops CF representations of warm items for robust training, which encourages robust interacation predictions over cold items solely based on feature representations. MTPR~\cite{du2020learn} instead replaces CF representations of warm items with zero vectors and applies multi-task learning for robust training.
% For example, ~\cite{volkovs2017dropoutnet} randomly drops CF representations of warm items for robust training. 
% and~\cite{du2020learn} replaces CF representations of warm items with zero vectors and applies multi-task learning for robust training. 
% Instead of dropping CF representations, MTPR~\cite{du2020learn} uses a method of replacing warm item CF representations with zero vectors and applies multi-task learning for more robust training. 
2) Auxiliary loss-based methods introduce different auxiliary losses for minimizing the distance between the feature and CF representations~\cite{wei2021contrastive,chen2022generative}. 
% For example,~\cite{wei2021contrastive} maximizes the mutual information between feature and CF representations spaces via contrastive loss, and~\cite{chen2022generative} utilizes adversarial training to effectively bridge the gap between feature and CF spaces.
% However, previous methods do not take into account the problem of temporal item feature shifts. 
% When the feature extractor is trained on warm items using ERM training, it will achieve a minimized expected loss on the same distribution as the training set, with the majority group in warm items being well-learned. 
% Typically, prior works attempt to achieve minimized expected loss on the same distribution as the warm items, where it well captures the feature representations of the majority warm item group yet under-represents other groups. 
% Nonetheless, the item features are shifting temporally from warm to cold items, where the majority groups of cold items can be different from the warm ones, resulting in poor generalization of feature extractor over cold items. 
However, previous methods suffer from temporal feature shifts from warm to cold items. To solve this issue, a concurrent study~\cite{wang2023equivariant} explores equivariant learning over minority groups of warm items. Differently, we leverage the shifting trend and emphasize the optimization of the potentially popular item groups. 
% In this work, we highlight the importance of strengthening generalization ability of feature extractor against temporal item feature shifts, where the key lies in considering the shifting trend of item features. 
% the temporally shifted distributions that reflect the 
% temporally shifted distributions that reflect the 
% into consideration and propose the goal of improving the generalization ability of cold-start recommender models by optimizing performance on possible shifted distributions to achieve robust prediction over cold items.

% \subsection{Distributionally Robust Optimization}
\noindent$\bullet\quad${\textbf{Distributionally Robust Optimization.}} 
% 参考related work共享文档 
% 先讲dro的motivation。然后1～2句话概括一下dro的general做法是什么（可以参考introduction p4）。
% 然后可以分类讨论一下dro的历史发展工作，从f-divergence,到Wasserstein distance，再到group dro，再到parametric dro等。和上面那个subsection一样，每种方法用1~2句话概括核心思想以及具体是怎么做的。然后对于每一类方法可以加一个例子，可以参考之前的dro 文章看看咋写
%  However，直接把dro用到cold start recommendation会suffer from inconsistency issue（这个回头可以参考下introdiction p4写法）。由于dro会overemphasize minority group which might not be necessarily popular in future，并且overemphasis会牺牲其他group的performance，包括那些可能会在未来变得popular的group。
% 为了解决这个问题，我们提出要利用推荐系统中的temporal shift来捕捉shifting trend的信号，并且用来调整dro对不同group的关注度，从而提升feature extractor的泛化能力，同时保证tail performance以及很可能变得popular的item group。
% 再讨论下dro和其他方法的区别，参考related work 共享文档
% DRO is an effective approach that aims to achieve uniform performance against distribution shifts by optimizing the worst-case performance over a pre-defined uncertainty set~\cite{rahimian2019distributionally,michel2022distributionally}. 
DRO aims to achieve uniform performance against distribution shifts~\cite{he2022causpref} by optimizing the worst-case performance over a pre-defined uncertainty set~\cite{rahimian2019distributionally,michel2022distributionally}. %,sagawa2019distributionally
% The exploration of DRO methods is primarily based on the definition of the uncertainty set. 
The most representative line of work is discrepancy-based DRO which defines the uncertainty set as a ball surrounding the training distributions with different discrepancy metrics~\cite{duchi2018learning,staib2019distributionally,liu2022distributionally}. 
% For example, DORO~\cite{zhai2021doro} proposes a refined risk function based on the Cressie-Read family uncertainty set to avoid the "hardest" instances, which may be outliers. 
% For example, DORO~\cite{zhai2021doro} leverages the Cressie-Read family of R$\Acute{\text{e}}$nyi divergence to define the uncertainty set, and GDRO~\cite{liu2022distributionally} is developed based on Wasserstein Distance.
% For example,~\cite{zhai2021doro} leverages the Cressie-Read family of R$\Acute{\text{e}}$nyi divergence to define the uncertainty set, and~\cite{liu2022distributionally} is developed based on Wasserstein Distance.
% Due to the fact that discrepancy-based DRO unnecessarily considers implausible distributions (\ie over-pessimism issue~\cite{oren2019distributionally,sagawa2020distributionally,duchi2023distributionally}), 
Since discrepancy-based DRO suffers from over-pessimism issue~\cite{oren2019distributionally,sagawa2020distributionally,duchi2023distributionally}), 
% suffers from the over-pessimism issue, where implausible distributions are unnecessarily considered, 
another line of research falls into Group-DRO~\cite{zhou2021examining,goel2020model}. It defines the uncertainty set as a set of mixtures of subgroups in the training set, encouraging DRO to focus on meaningful distribution shifts~\cite{oren2019distributionally,wen2022distributionally}. 
Some prior work~\cite{zhou2023distributionally} explores DRO to alleviate long-tail users and items for warm-start recommendation, \eg S-DRO~\cite{wen2022distributionally}, PDRO~\cite{zhao2023popularity}. 
However, directly applying DRO to cold-start recommendation may cause inconsistency issue. 
% The overemphasis on the minority group of warm items may weaken the expressiveness of the groups that are likely to become popular in the upcoming cold items. 
In this work, we consider leveraging a temporally DRO to focus on the mitigation of temporal item feature shifts for cold-start recommendation. 
% , thus limiting the robustness enhancement for cold item recommendation. 
% In this work, we propose to leverage the temporal shifting trend to guide DRO to improve the generalization ability of the feature extractor against temporal item feature shifts. 
% Besides, potential solution for  compared and discussed in Section~\ref{sec:experiment}.  

% **Appendix
More detailed discussions on related works are presented in Appendix~\ref{appendix:related_work}, including other literature on robustness enhancement, such as invariant learning~\cite{arjovsky2019invariant,du2022invariant,koyama2020out,liu2021heterogeneous,pan2023discriminative} and re-weighting strategy~\cite{zhang2021causal,kim2021premere,yang2023dgrec}. 

% Other literature on robustness has also been extensively studied. 
% Invariant learning~\cite{arjovsky2019invariant,du2022invariant,koyama2020out,liu2021heterogeneous} considers the invariant part robust to distribution shifts but it overlooks the variant part, which may be essential for prediction. 
% Re-weighting loss~\cite{zhang2021causal,kim2021premere,yang2023dgrec} assigns weights to samples, which however relies heavily on correlations between group density and task difficulty and yields inferior performance than DRO (\cf Table 1 in ~\cite{wen2022distributionally}).

\section{Preliminary}
\label{sec:preliminaries}
% 参考ICLR’20 ON THE IMPORTANCE OF
% 宏观角度讲模型优化过程，input，output，model parameter等的符号定义
% ERM是怎么做的，公式，但是这么做依赖于IID的假设，并且会容易拟合到占比较多的数据上。
% DRO的motivation，DRO的通用公式
% traditonal DRO的问题 （group DRO的motivation），group dro的公式，group dro的作用。
% In this section, we first retrospect the learning of the cold-start item recommender model and reveal its vulnerability against temporal item feature shift. Then we introduce the background of DRO and uncover its potential as well as limitations for cold-start item recommendation. 

% In this section, we first retrospect the learning of the cold-start item recommender model and reveal its vulnerability against temporal item feature shifts. Then we delve into the background of DRO and uncover its potential as well as limitations for cold-start item recommendation. 

% 介绍现有的cold-start recommender是怎么做的 
% \vspace{3pt}
% \subsection{Cold-start Recommendation}
\subsubsection{Cold-start Recommendation.}
% \noindent\textbf{$\bullet $ Cold-start Recommendation.} 
% ERM 是不是就直接写cold start recommendation里的loss 是类似ERM
% 参考equivariant learning那篇和其他cold-start文章，考虑是否区分implicit和explicit。
% 先定义所有要用到的notation, 给定user, warm item, warm item feature, cold item, cold item feature, user-item interactions，cold-start recommender model aims to learn a feature extractor XX and a interaction predictor XX for warm and cold-start recommendation. 

% To address the cold-start item issue, the cold-start item recommender models utilize the item features (\eg categories and visual features) to predict the user-item interactions. 
% Specifically, given the users $\mathcal{U}$, warm items $\mathcal{I}_w$ with features $\{\bm{s}_i|i\in \mathcal{I}_w\}$, cold items $\mathcal{I}_c$ with features $\{\bm{s}_i|i\in \mathcal{I}_c\}$, and user-item interactions $\mathcal{D}=\{(u,i,y_{ui})|u\in\mathcal{U}, i\in\mathcal{I}_w\}$ 
% % , where $y_{ui}\in\{0,1\}$ implies whether the user $u$ likes the item $i$ ($y_{ui}=1$) or not ($y_{ui}=0$), 
% with $y_{ui}\in\{0,1\}$ indicating whether the user $u$ likes the item $i$ ($y_{ui}=1$) or not ($y_{ui}=0$), 
% the cold-start recommender model aims to learn a feature extractor $q_{\theta_1}(\cdot)$, an interaction predictor $f_{\theta_2}(\cdot)$, and the CF representations of users and items $\{\bm{z}_u,\bm{z}_i\}$ for warm and cold item recommendation. 
To address the cold-start item issue, existing methods leverage the item features (\eg categories and visual features) to predict the user-item interactions. 
% todo: 一句改为两句
% Specifically, given the users $\mathcal{U}$, warm items $\mathcal{I}_w$ with features $\{\bm{s}_i|i\in \mathcal{I}_w\}$, cold items $\mathcal{I}_c$ with features $\{\bm{s}_i|i\in \mathcal{I}_c\}$, and user-item interactions $\mathcal{D}=\{(u,i,y_{ui})|u\in\mathcal{U}, i\in\mathcal{I}_w\}$ 
Specifically, given the users $\mathcal{U}$, warm items $\mathcal{I}_w$ with features $\{\bm{s}_i|i\in \mathcal{I}_w\}$, 
and user-item interactions $\mathcal{D}=\{(u, i, y_{ui})|u\in\mathcal{U}, i\in\mathcal{I}_w\}$ 
% , where $y_{ui}\in\{0,1\}$ implies whether the user $u$ likes the item $i$ ($y_{ui}=1$) or not ($y_{ui}=0$), 
with $y_{ui}\in\{0,1\}$ indicating whether the user $u$ likes the item $i$ ($y_{ui}=1$) or not ($y_{ui}=0$), 
% the cold-start recommender model aims to learn a feature extractor, an interaction predictor, and the CF representations of users and items for aligning feature representations with user-item interactions. 
the cold-start recommender model aims to learn a feature extractor, an interaction predictor, and the CF representations of users and items for aligning feature representations with user-item interactions. 
% We denote the learnable parameters of cold-start recommender as $\theta$. 
% for both warm and cold item recommendation. 
% The learnable parameters of feature extractor and the interaction predictor are optimized by Empirical Risk Minimization (ERM), where the risk function can be BPR loss or BCE loss.
% Formally, we have
% \theta=miminize XXX 这里用上recommender的notation，ERM的公式
The learnable parameters of the cold-start recommender model, 
% (\ie feature extractor, interaction predictor, and CF representations), 
% denoted as $\theta=\{\theta_1,\theta_2, \bm{z}_u, \bm{z}_i\}$, 
denoted as $\theta$, 
are optimized via Empirical Risk Minimization (ERM). Formally, we have
\begin{equation}
\small
\label{eqn:erm_optimization}
\begin{aligned}
% & \mathcal{L} = {\mathcal{L}_{r} + \mathcal{L}_{a}} + \epsilon\cdot\mathcal{L}_{1}+\mathcal{L}_{2}+\mathcal{L}_{3},
& \theta^*_{ERM} := \mathop{\arg\min}_{\theta\in\Theta} \mathbb{E}_{(u,i,y_{ui})\in\mathcal{D}} [\mathcal{L}\left(\theta;(u,i, y_{ui},\bm{s}_i)\right)], \\
% & \mathop{\arg\min}_{\theta,\phi,\{\bm{z}_u\},\{\bm{z}_i\}} \mathbb{E}_{(u,i,y_{ui})\in \mathcal{D}}\mathcal{L}_{r}\left(f_\theta(\bm{z}_u, \bm{z}_i), y_{ui}\right) + \mathcal{L}_{a}\left(g_\phi(\bm{c}_i), \bm{z}_i\right),\\
\end{aligned}
\end{equation}
% \vspace{3pt}
where $\mathcal{L}(\cdot)$ is the loss function of the cold-start recommender model and is particularly tailored to different cold-start methods to regulate the alignment. 
Nevertheless, such a learning paradigm merely minimizes the expected loss under the same distribution as the training data~\cite{rahimian2019distributionally}. 
% This results in the learned feature extractor that encodes the feature of the majority group of warm items more accurately but under-represent the minority groups. 
% This results in the learned feature extractor that encodes feature representations more accurately for the majority group in warm items but under-represent the minority groups, which may become popular in cold items. 
The feature extractor could under-represent the minority groups~\cite{wen2022distributionally}, which however might be popular in cold items, leading to the vulnerability to 
the shifted cold item features. 
% the temporally shifted distributions of cold items. 
% As such, 
% in the realm of item feature shifts, 
% the learned feature extractor is vulnerable to the temporally shifted distributions of cold items, leading to poor generalization ability.

% \vspace{3pt}
\subsubsection{Distributionally Robust Optimization.}

% DRO, 
To alleviate temporal feature shifts, DRO\footnote{We adopt Group-DRO to avoid over-pessimism issue (refer to Appendix~\ref{appendix:DRO} for details).} 
% \footnote{We adopt Group-DRO to avoid over-pessimism issue.} 
% \footnote{}
is an effective solution that could achieve consistently high performance across various distribution shifts~\cite{zhou2021examining,duchi2018learning,oren2019distributionally,sagawa2020distributionally,hu2018does}. 
In detail, DRO assumes the training distribution to be a mixture of $K$ pre-defined groups $\{P_i|i=1,\dots,K\}$. 
% \ie $\mathcal{P}:=\sum_{i=1}^{K}w_iP_i$, where $w_i$ is the mixture ratio of the group $i$. 
% 
Then, it optimizes the worst-case performance over the $K$ subgroups for controlling the performance lower bound. Formally, 
% \begin{equation}
% \small
% \label{eqn:dro_optimization}
% \begin{aligned}
% & \theta^*_{\text{DRO}} := \mathop{\arg\min}_{\theta\in\Theta}
% \left\{\mathop{\sup}_{Q\in\mathcal{Q}}\mathbb{E}_{(u,i,y_{ui})\in{Q}} [\mathcal{L}\left(\theta;(u,i, y_{ui},\bm{s}_i)\right)]\right\}. \\
% \end{aligned}
% \end{equation}
% where the uncertainty set $\mathcal{Q}$ is typically defined as a ball surrounding the training distributions~\cite{duchi2018learning,liu2022distributionally}, which however might unnecessarily consider implausible distributions~\cite{liu2022distributionally,oren2019distributionally} (refer to Appendix~\ref{appendix:DRO} for details). 
% To overcome this issue, Group-DRO is proposed to find the potential shifted distributions on a group level~\cite{oren2019distributionally,sagawa2020distributionally,hu2018does}. 
% In Group-DRO, the training distribution $\mathcal{P}$ is assumed to be a mixture of $K$ pre-defined groups 
% $\{P_i|i=1,\dots,K\}$, \ie $\mathcal{P}:=\sum_{i=1}^{K}w_iP_i$, where $w_i$ is the mixture ratio of the group $i$. 
% where the uncertainty set $\mathcal{Q}:=\{\sum_{i=1}^{K}w'_iP_i: w'_i\in\Delta_K\}$, and $\Delta_K$ is the $K-1$-dimensional probability simplex. Since the maximum of $\sum_{i=1}^{K}w'_iP_i$ is attained at a vertex due to linearity, the objective of Group-DRO is reformulated as 
\begin{equation}
\small
\label{eqn:gdro_optimization}
\begin{aligned}
% & \theta^*_{\text{Group-DRO}} := \mathop{\arg\min}_{\theta\in\Theta} \
% \left\{\hat{\mathcal{R}} (\theta):=\mathop{\max}_{j\in\{1,\dots,K\}}\mathbb{E}_{(u,i,y_{ui})\sim P_j} [\mathcal{L}\left(\theta;(u,i, y_{ui},\bm{s}_i)\right)]\right\}. \\
\theta^*_{\text{DRO}} := \mathop{\arg\min}_{\theta\in\Theta} \left\{
\mathop{\max}_{j\in[K]}\mathbb{E}_{(u,i,y_{ui})\sim P_j} [\mathcal{L}\left(\theta;(u,i, y_{ui},\bm{s}_i)\right)]
\right\}. 
\end{aligned}
\end{equation}
% \vspace{3pt}
A practical solution to
Eq. (\ref{eqn:gdro_optimization}) is to conduct interleave step-wise optimization~\cite{piratla2022focus,sagawa2020distributionally}. 
% where the learnable parameters $\theta$ are updated based on the group with the highest empirical loss. 
Specifically, at each update step $t$, DRO first selects the group with the worst empirical performance:
\begin{equation}
\small
\label{eqn:gdro_group_selection}
% \left\{
\begin{aligned}
j^*&=\mathop{\arg\max}_{j\in\{1,\dots,K\}} \mathbb{E}_{(u,i,y_{ui})\sim {P}_j}[\mathcal{L}(\theta;(u,i,y_{ui},\bm{s}_i))] \ \\
&\approx\mathop{\arg\min}_{j\in\{1,\dots,K\}} -\Bar{\mathcal{L}}_{j}, \\
\end{aligned}
% \right.
\end{equation}
where $ \Bar{\mathcal{L}}_{j} = \frac{1}{N_j}\sum\nolimits_{(u,i,y_{ui})\sim \Tilde{P}_j}\mathcal{L}_j(\theta;(u,i,y_{ui},\bm{s}_i))$,
$\Tilde{P}_j$ is the empirical distribution of group $j$ in dataset $\mathcal{D}$, and $N_j$ is the number of samples in group $j$. 
% Subsequently, based on the selected group, the model parameters $\theta$ are updated via gradient descent, \ie $\theta^{t+1}=\theta^t-\eta\nabla_\theta \Bar{\mathcal{L}}_{j^*}(\theta^t)$, where $\eta$ is the learning rate. 
Subsequently, the model parameters $\theta$ are updated based on the selected group, \ie $\theta^{t+1}=\theta^t-\eta\nabla_\theta \Bar{\mathcal{L}}_{j^*}(\theta^t)$, where $\eta$ is the learning rate. 

% \vspace{2pt}
Despite the success of DRO in various domains (\eg image classification~\cite{zhai2021doro,sagawa2020distributionally}, natural language modeling~\cite{oren2019distributionally,michel2022distributionally}), directly applying DRO in cold-start recommendation faces an inconsistency issue. 
% To be precise, in the context of temporal item feature shifts, the majority and minority item groups in the training set may not necessarily remain to be the majority and the minority in the test set. 
It is likely that DRO will overemphasize the minority group in warm items at the expense of performance of other groups~\cite{wen2022distributionally}. Besides, the majority and minority item groups may change due to temporal feature shifts, thereby hurting the cold item performance (\cf Section~\ref{sec:introduction}).  
% Traditional DRO tends to prioritize minority groups in warm items at the expense of other groups' performance. 
% Since minority groups in warm items may not ensure their popularity in subsequent cold items, excessive emphasis on these groups could inadvertently diminish the performance of popular groups in cold items.
% todo： 容易fit到minority，考虑提供emipirical resutls吗？

\begin{figure*}[t]
\centering
\includegraphics[scale=0.9]{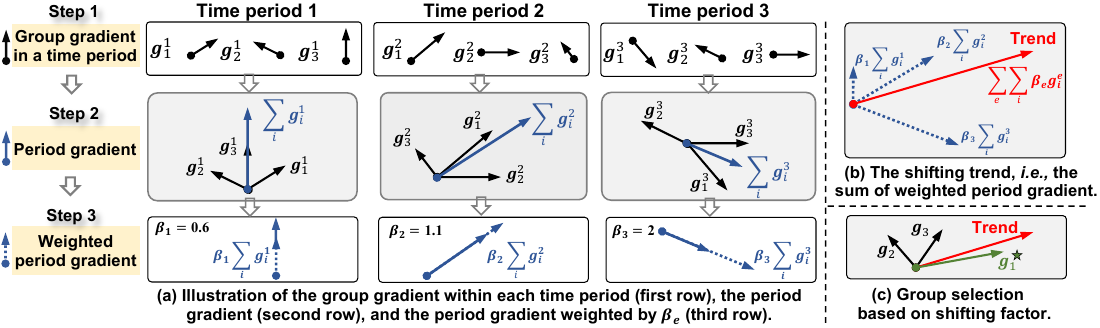}
% \caption{Illustration of shifting factor with three groups and three time periods. The three subfigures on the left of (a) depict the group gradient in each time period and the period gradient (\ie the blue vectors) is depicted on the right of (a). (b) depicts the shifting-ware holistic gradients over the temporally shifted distributions (\ie the black vector) in the upper subfigure, and the group gradient at the bottom. (c) demonstrates the overall shifting factor and the red vector is the constant vector indicating the shifting trend.}
% \caption{Illustration of shifting factor with three groups and three time periods (\ie $i\in\{1,2,3\}$ and $e\in\{1,2,3\}$). (a) depicts each group gradient within a time period (first row), the period gradient (second row), and the period gradient weighted by $\beta_e$ (third row). (b) depicts the shifting-aware holistic gradient over the temporally shifted distributions. (c) demonstrates the selected group (marked in green) based on the similarity between the group gradient and the shifting trend.}
\caption{Illustration of the shifting factor with three groups and three time periods (\ie $i\in\{1,2,3\}$ and $e\in\{1,2,3\}$). (a) depicts the three steps of obtaining the weighted period gradient in each time period. And then, by summing up the weighted period gradient, we can obtain the shifting trend as shown in (b). Finally, the shifting factor for each group is obtained by calculating the similarity between the group gradient and the shifting trend as presented in (c).}
\label{fig:methods}
\end{figure*}

\section{Temporally DRO}
\label{sec:method}

% 要不要先讲一下整个方法的motivation，介绍一下objective然后引出理论上dro要优化XXX。在介绍具体的实现。也就是说目标优化loss先顺着motivation和objective引出来，然后再写实现这个loss要做啥。

% 为了避免traditionall dro overemphasize 了minority group 会导致inconsistency的问题，牺牲了很多本不应该牺牲的item expressiveness，由于推荐系统自然存在temporal shift，因此我们可以利用好这个shifting trend来guide DRO。 于是我们提出two objective for dro training: 1) xxx 2) xxx (cf introduction)
% In order to alleviate the overemphasis on minority groups and strengthen the expressiveness of item groups that should not be compromised, the key lies in considering the temporal shifting trend in recommender systems to guide DRO. 
% To enhance the expressiveness of item groups that should not be compromised, 
To alleviate the impact of temporal feature shifts for cold-start recommendation, 
% and mitigate the overemphasis on minority groups, 
% it is essential to consider the temporal shifting trend as a guiding principle for DRO. 
% todo: two objectives 和introduction的两条保持一致 - ok
% To achieve this, 
we propose two new objectives for DRO training: 
1) enhancing the worst-case optimization on minority groups to raise the lower bound of performance, and 
2) capturing the temporal shifting trend of item features and emphasizing the optimization of groups that are likely to become popular. 
% 1) optimize the worst-case performance of the minority group for tail performance guarantee, and 2) capture the temporal shifting trend for improving the groups that are likely to become popular. 
% In the following, we will present our method based on the setting of Group-DRO due to its high practicability and efficiency~\cite{sagawa2020distributionally}. 
% To implement our method, we adopt the Group-DRO setting, which is highly practical and efficient~\cite{sagawa2020distributionally,oren2019distributionally}. 

\subsection{Group Selection}
% It is noted that worst-case group selection in DRO (Eq. (\ref{eqn:gdro_group_selection})) is critical in strengthening the model's robustness~\cite{piratla2022focus}. As such, we propose a novel temporal DRO called Temporal DRO (TDRO), which introduces two factors in group selection: 
It is noted that the group selection plays a critical role in DRO (Eq. (\ref{eqn:gdro_group_selection})) to strengthen the model's robustness~\cite{piratla2022focus}. 
As such, we propose a novel TDRO, which introduces two factors in group selection: 
1) \textit{{a worst-case factor}} to focus more on minority groups with larger losses and
% gives them priorities for performance optimization, 
give them priorities for group selection, 
and
% Besides, to improve the performance of the groups that are likely to be popular in the future, we introduce 2) {\textit{shifting trend factor}} to capture the temporal shifting trend and alleviate the unnecessary sacrifice on these groups.
% 2) {\textit{Shifting factor.}} 
% To alleviate the overemphasis on one particular worst-case group, we consider enhancing performance among all groups and leveraging the temporal shifting trend to emphasize the potentially popular groups in cold items. 
2) {\textit{a shifting factor}} to emphasize the potentially popular groups in cold items by leveraging the temporal shifting trend. Besides, the shifting factor can alleviate the overemphasis on one particular worst-case group. 
% Intuitively, the guiding principle of the shifting factor for group selection is: train on the group that can enhance performance among all groups in each time period, and 
% To alleviate the unnecessary sacrifice on groups likely to become popular in cold items, two key considerations
% 为了照顾minority group，我们可以关注那些performance比较差的（loss比较大的）group，给他们更高的优先级进行performance optimization. 而关于如何挖掘shifting trend 的信号，我们可以利用推荐系统自然存在的supervise signal (temporal shift), inject到dro的训练中。
% Specifically, 我们首先对warm item通过item feature来划分成不同的group, 我们先把interaction按照时间排序，然后划分进不同的时间段作为一个环境e，xxx
% 我们希望在考虑group-level performance的同时，我们还希望，更新这个group之后，能让其他temporally shifted distribution上的sample能尽可能多的提高performance，而shifting trend 则体现在约靠近test时间的time period越重要。formally,

% 公式 -> 化简 -> gradient 的公式。
% 介绍第一个公式, 就是我们希望所有temporally shifted distrubtions in various time periods中的performance都能有很大的提升，并且希望越贴近shifting trend的越重要。
% 
% 参考代码推导_v2的pdf

% \vspace{3pt}
% \noindent\textbf{$\bullet$ {\textbf{Shift-guided group selection.}}}
\subsubsection{\textbf{Shifting Trend-guided Group Selection.}} 
% Typically, the shifting trend is characterized by the gradual shifts from one time period to another (see Figure~\ref{fig:intro_shift}), and  
% the trend towards potentially popular items is getting more significant as time periods draw closer to the test phase~\cite{wang2023causal}. 
% Therefore, we consider the shifting factor to define a rule for group selection. That is we want to train on the group that yields the highest performance gains across all groups, while giving more emphasis to the time periods closer to the test phase. 

In detail, we first split the warm items into $K$ groups via $K$-means clustering based on their item features (\eg visual features of thumbnails). We then split the chronologically sorted interactions into $E$ time periods, $e\in\{1,\dots,E\}$. 
% , where each time period contains the interactions during a certain time period. 
We denote 
% the average loss of group $j$ in all time periods as $\Bar{\mathcal{L}}_j(\cdot)$, and 
the average loss of group $i$ in time period $e$ as $\mathcal{L}_i^e(\cdot)$. 
At each update step $t$, we consider both the worst-case factor and the shifting factor to select the group $j^*$ for optimization, which is formulated as
% 这里放公式
% 搞一个朝下的话括号，标出两个factor
% 并解释每个符号的意义 \underbrace{\mathbb{E}_{q(\bm{x}_1|\bm{x}_0)}\left[\log p_\theta(\bm{x}_0|\bm{x}_1)\right]}_{\small(\text{reconstruction term})}
% \begin{multline}
% \small
% \label{eqn:tdro_group_selection_1}
% j^*=\mathop{\arg\min}_{j\in\{1,\dots,K\}} \underbrace{-(1-\lambda)\Bar{\mathcal{L}}_j(\theta^t)}_{(\text{worst-case factor})} \\
% + \underbrace{\lambda\sum_{e=1}^{E}\sum_{i=1}^{K}\beta_e \mathcal{L}_i^e(\theta^t-\eta\nabla_\theta \Bar{\mathcal{L}}_j(\theta^t))}_{(\text{shifting factor})}, 
% \end{multline}
\begin{equation}
\small
\label{eqn:tdro_group_selection_1}
\begin{aligned}
j^*=\mathop{\arg\min}_{j\in\{1,\dots,K\}}& \underbrace{-(1-\lambda)\Bar{\mathcal{L}}_j(\theta^t)}_{\let\scriptstyle\textstyle
    \substack{(\text{worst-case factor})}} \\
+ &\underbrace{
      \lambda\sum_{e=1}^{E}\sum_{i=1}^{K}\beta_e \mathcal{L}_i^e(\theta^t-\eta\nabla_\theta \Bar{\mathcal{L}}_j(\theta^t))}_{\let\scriptstyle\textstyle
    \substack{(\text{shifting factor})}},
\end{aligned}
\end{equation}
% where $\Bar{\mathcal{L}}_j(\cdot)$ denotes the average loss of group $j$ in all time periods, and 
where $\lambda$ is the hyper-parameter to balance the strength between two factors. 
% It is highlighted that:
% \begin{itemize}[leftmargin=*]
%     \item The \textbf{\textit{worst-case factor}} is designed to consider the group performance and prioritize the groups with larger loss values. As such, TDRO can raise the lower bound of performance, similar to DRO~\cite{sagawa2020distributionally}.
%     % todo: 这里和introduction p5对应一下
%     \item The \textbf{\textit{shifting factor}} consists of two perspectives. 1) To alleviate the overemphasis on one particular worst-case group, it considers performance improvements for \textit{all} groups when selecting the group for optimization. 
%     2) To emphasize the potentially popular groups in cold items, it leverages the temporal shifting trend by casting more emphasis on the time periods closer to the test phase during group selection. 
%     \todo{(}This is because the temporal shifting trend is characterized by the gradual shifts from one time period to another (see Figure~\ref{fig:intro_shift}), and the trend towards potentially popular items is getting more significant as time periods draw closer to the test phase~\cite{wang2023causal}. \todo{)}
%     % In each time period, the potential benefits of optimizing the model on the selected group to optimize the performance for all groups. Moreover, the monotonically increasing $\beta_e$ casts more emphasis on the time periods closer to the test phase, pushing TDRO to improve the performance of the groups that are likely to become popular in cold items.  
%     % Further explanations are presented in Section ~\ref{subsubsec:interpretation}.
% \end{itemize}
The \textbf{\textit{worst-case factor}} calculates the loss value of each group $\Bar{\mathcal{L}}_j(\theta^t)$ for group selection. The group with a larger loss will have a smaller $-\Bar{\mathcal{L}}_j(\theta^t)$, thus being more likely to be selected. 
Besides, the \textbf{\textit{shifting factor}} consists of two perspectives: 
\begin{itemize}[leftmargin=*]
    \item To alleviate the overemphasis on one particular worst-case group, the shifting factor selects the optimization group to improve the performance on \textit{all} groups. 
    % considers performance improvements for \textit{all} groups when selecting the group for optimization. 
    Specifically,
    $\theta^t-\eta\nabla_\theta \Bar{\mathcal{L}}_j(\theta^t)$ is the updated parameters if we choose group $j$ for optimization. Thereafter, the loss of each group $i$ in a time period $e$ after parameter updating will be $\mathcal{L}_i^e(\theta^t-\eta\nabla_\theta \Bar{\mathcal{L}}_j(\theta^t))$. And the performance improvements for all groups across all periods are measured by $\sum_{e=1}^{E}\sum_{i=1}^{K} \mathcal{L}_i^e(\theta^t-\eta\nabla_\theta \Bar{\mathcal{L}}_j(\theta^t))$. 
    \item To emphasize the potentially popular groups in cold items, the shifting factor upweights the later time periods closer to the test phase. In detail, we use $\beta_e$ to re-weight the performance improvements over all groups for each time period $e$. We define $\beta_e=\exp(p\cdot e)$, where a later period $e$ will have a higher weight and $p>0$ is the hyper-parameter to control the steepness. A smaller $p$ encourages time periods to be uniformly important, while a larger $p$ upweights the time periods closer to the test phase. 
    % Therefore, the shifting factor in Eq. (\ref{eqn:tdro_group_selection_1}) pushes TDRO 
% Therefore, the shifting factor in Eq. (\ref{eqn:tdro_group_selection_1}) measures the performance improvements over all groups in time periods with different importance 
% $\beta_{e\in[E]}$. 

\end{itemize}

However, directly applying Eq. (\ref{eqn:tdro_group_selection_1}) for group selection will incur extensive resource costs as we need to consider all possible cases of the updated parameters. 
% to obtain the performance gains and select the optimal group. 
% thus limiting the applicability of TDRO, especially for large-scale cold-start item recommendation. 
% To lessen the computing burden and enable efficient group selection, 
% To achieve efficient group selection and facilitate TDRO for large-scale cold-start item recommendation, 
Fortunately, 
we can approximate Eq. (\ref{eqn:tdro_group_selection_1}) into a gradient-based formulation via First-order Taylor formulation (see Appendix~\ref{appendix:derivation} for detailed derivation):  
% **Appendix
% By applying First-order Taylor(see Appendix~\ref{sec:appendix} for detailed derivation), Eq. (\ref{eqn:tdro_group_selection_1}) can be approximated into a gradient-based formulation as, 
 % 一阶泰勒，（展开一个式子？），到gradient的形式，推导pdf (4)，
 % 放方法示意图，
% \subsubsection{Time-aware re-weighting strategy.}
\begin{equation}
\small
\label{eqn:tdro_group_selection_taylor}
\begin{aligned}
% & j^*=\mathop{\arg\max}_{j\in\{1,\dots,K\}} \underbrace{(1-\lambda)\Bar{\mathcal{L}}_j(\theta^t)}_{(\small\text{worst-case factor})} + \underbrace{\lambda {\bm{g}}_j^T\sum_{e=1}^E\sum_{i=1}^K {\beta}_e\bm{g}_i^e}_{(\small\text{shifting factor})}, \\
& j^*=\mathop{\arg\max}_{j\in\{1,\dots,K\}} \underbrace{(1-\lambda)\Bar{\mathcal{L}}_j(\theta^t)}_{\let\scriptstyle\textstyle
    \substack{(\text{worst-case factor})}} + \underbrace{\lambda {\langle\bm{g}}_j,\sum_{e=1}^E\sum_{i=1}^K {\beta}_e\bm{g}_i^e\rangle}_{\let\scriptstyle\textstyle
    \substack{(\text{shifting factor})}}, \\
\end{aligned}
\end{equation}
% where $\bm{g}_j=\nabla_\theta \Bar{\mathcal{L}}_j(\theta)$ denotes the gradient of average loss of group $j$ \wrt parameters $\theta$,
where $\bm{g}_j=\nabla_\theta \Bar{\mathcal{L}}_j(\theta)$ denotes the gradient of the average loss of group $j$, and
$\bm{g}_i^e=\nabla_\theta {\mathcal{L}}^e_i(\theta)$ denotes the gradient of group $i$'s average loss in time period $e$. The 
$\langle\cdot,\cdot\rangle$ represents the inner product computation. 
Since $\sum_{e=1}^{E}\sum_{i=1}^{K} {\beta}_e\bm{g}_i^e$ is a constant vector (referred to as \textit{shifting trend}) for any group $j$, we can avoid this cumbersome computations in Eq. (\ref{eqn:tdro_group_selection_1}) for efficient group selection. 
% Further interpretations of the gradient-based shifting factor are presented in the following. 

% \vspace{3pt}

% \vspace{3pt}
% \noindent\textbf{$\bullet$ Interpretation of shifting trend.} 
\subsubsection{\textbf{Interpretation of Shifting Factor.}}\label{subsubsec:interpretation} 

% todo: 画图shifting trend的示意图 - ok 

For an intuitive understanding of the gradient-based shifting factor, we visualize a toy example in Figure~\ref{fig:methods}, where we set $K=3$ and $E=3$.  
% where the numbers of time period and groups are both three. 
% where $K=3$ and $E=3$.
% where the items are divided into three groups and the samples into three time periods. 

% \vspace{2pt}
\noindent$\bullet\quad$\textbf{\textit{Factor decomposition.}} 
% We first illustrate the gradient of the group samples in each time period in Figure~\ref{fig:methods}(a), where we have three group-wise gradients for each time period $\bm{g}_i^e$ in Eq. (\ref{eqn:tdro_group_selection_taylor}), referred to as period gradients. 
% This corresponds to the term $\bm{g}_i^e$ of shifting factor in Eq. (\ref{eqn:tdro_group_selection_taylor}). 
% And then we could obtain period gradient $\sum\bm{g}_i^e$ by summing up the group gradient in each time period (see figures on the right in Figure~\ref{fig:methods}(a)). 
As shown in Figure~\ref{fig:methods}(a), we have three decomposed group gradients,  $\bm{g}_{i\in\{1,2,3\}}^e$, for each time period $e$.
We can then obtain the period gradient $\sum_{i=1}^{K}\bm{g}_i^{e}$ of time period $e$ by summing up the decomposed group gradients. 
Since the gradient indicates the optimization direction, the sum of the gradient within each time period, \ie period gradient, represents the optimal updating direction in each temporally shifted distribution. 
Subsequently, by multiplying the period importance $\beta_e$ to each time period and summing up the weighted period gradient, we can obtain the shifting trend $\sum_{e=1}^{E}\sum_{i=1}^K {\beta}_e\bm{g}_i^e$ that 
% considers the temporal shifting trend as shown in Figure~\ref{fig:methods}(b).
reflects optimization direction on potentially popular groups (Figure~\ref{fig:methods}(b)).

\noindent$\bullet\quad$\textbf{\textit{Factor interpretation.}} 
Finally, the shifting factor is obtained by calculating the inner product of the shifting trend and the group gradient $\bm{g}_j$ (see Figure \ref{fig:methods}(c)). Since the shifting trend is a constant vector for all groups, the shifting factor essentially measures the similarity between each group gradient and the shifting trend, \ie optimization direction emphasizing the potentially popular item groups. 
% It can also be interpreted as the benefits of optimizing the model over group $j$ in terms of the performance enhancement
% over weighted temporally shifted distributions.
% over temporally shifted distribution reflecting the shifting trend. 

% \vspace{3pt}
As for model optimization at each step, we first select the optimal group $j^*$ via Eq. (\ref{eqn:tdro_group_selection_taylor}), and then update the parameters $\theta$ by gradient descent $\theta^{t+1}=\theta^t-\eta\nabla_\theta \Bar{\mathcal{L}}_{j^*}(\theta^t)$.
\subsection{Gradient Smoothing}
Despite the success of step-wise optimization in many applications~\cite{sagawa2020distributionally}, directly employing such strategy in recommender systems 
% with worst-case performance and large potential benefits 
% over the temporally shifted distributions 
suffers from training instability~\cite{wen2022distributionally}. 
% (see Figure~\ref{} for empirical evidence). 
As such, we follow the previous work~\cite{piratla2022focus,wen2022distributionally} by incorporating gradient smoothing for optimization from two aspects: group importance smoothing and loss consistency enhancement.

% \vspace{3pt}
\noindent\textbf{$\bullet$ {Group importance smoothing.}}
We consider assigning weight vector $\bm{w}$ for groups and regulate the weight dynamic by $\eta_w$. Formally,
% \begin{equation}
% \label{eqn:tdro_group_weight}
% \resizebox{1\hsize}{!}{$
% \left\{
% \begin{aligned}
% & \bm{w}^{t+1}=\mathop{\arg\max}_{{w_{i\in\{1,\dots,K\}}}\in\Delta_{K}} \sum_{i}w_i[(1-\lambda)\Bar{\mathcal{L}}_i(\theta)+\lambda\langle \bm{g}_i,\sum_{e=1}^E\sum_{j=1}^K {\beta_e}\bm{g}_j^e\rangle]-\frac{1}{\eta_w}\text{KL}(\bm{w},\bm{w}^t), \\
% &\theta^{t+1}=\theta^t-\eta\sum_i w_i^{t+1}\Bar{\mathcal{L}}_i(\theta^t), \\
% \end{aligned}
% \right.$}
% \end{equation}
\begin{equation}
\label{eqn:tdro_group_weight}
\resizebox{1\hsize}{!}{$
\begin{aligned}
% & \bm{w}^{t+1}=\mathop{\arg\max}_{{w_{i\in[K]}}\in\Delta_{K}} \sum_{i}w_i[(1-\lambda)\Bar{\mathcal{L}}_i(\theta)+\lambda\langle \bm{g}_i,\sum_{e=1}^E\sum_{j=1}^K {\beta_e}\bm{g}_j^e\rangle]-\frac{1}{\eta_w}\text{KL}(\bm{w},\bm{w}^t), \\
& \bm{w}^{t+1}=\mathop{\arg\max}_{{w_{i\in[K]}}} \sum_{i}w_i[(1-\lambda)\Bar{\mathcal{L}}_i(\theta)+\lambda\langle \bm{g}_i,\sum_{e=1}^E\sum_{j=1}^K {\beta_e}\bm{g}_j^e\rangle]-\frac{1}{\eta_w}\text{KL}(\bm{w},\bm{w}^t), \\
\end{aligned}
$}
\end{equation}
where $w_i$ is the $i$-th entry of $\bm{w}$, $\eta$ is the learning rate, and $\text{KL}(\bm{p},\bm{q})=\sum_ip_i\log\frac{p_i}{q_i}$ is the KL-divergence between $\bm{p}$ and $\bm{q}$. 
By applying KKT conditions (refer to Appendix~\ref{appendix:derivation}), we obtain the closed-form solution of Eq. (\ref{eqn:tdro_group_weight}): 
% **Appendix
% By applying KKT conditions (refer to Appendix~\ref{appendix:derivation}), we obtain the closed-form solution of Eq. (\ref{eqn:tdro_group_weight}): 
% \begin{equation}
% \label{eqn:tdro_group_weight_solution}
% \resizebox{1\hsize}{!}{$
% \left\{
% \begin{aligned}
% & w_i^{t+1}=\frac{w_i^t \exp(\eta_w[(1-\lambda)\Bar{\mathcal{L}_i}(\theta^t)+\lambda\langle \bm{g}_i,\sum_{e=1}^E\sum_{j=1}^K\beta_e\bm{g}_j^e\rangle])} {\sum_s w_s^t \exp(\eta_w[(1-\lambda)\Bar{\mathcal{L}_s}(\theta^t)+\lambda \langle \bm{g}_s,\sum_{e=1}^E\sum_{j=1}^K\beta_e\bm{g}_j^e\rangle])}, \\
% &\theta^{t+1}=\theta^t-\eta\sum_i w_i^{t+1}\Bar{\mathcal{L}}_i(\theta^t). \\
% \end{aligned}
% \right.$}
% \end{equation}
\begin{equation}
\label{eqn:tdro_group_weight_solution}
\resizebox{1\hsize}{!}{$
\begin{aligned}
& w_i^{t+1}=\frac{w_i^t \exp(\eta_w[(1-\lambda)\Bar{\mathcal{L}_i}(\theta^t)+\lambda\langle \bm{g}_i,\sum_{e=1}^E\sum_{j=1}^K\beta_e\bm{g}_j^e\rangle])} {\sum_s w_s^t \exp(\eta_w[(1-\lambda)\Bar{\mathcal{L}_s}(\theta^t)+\lambda \langle \bm{g}_s,\sum_{e=1}^E\sum_{j=1}^K\beta_e\bm{g}_j^e\rangle])}. \\
\end{aligned}
$}
\end{equation}
Thereafter, the model parameters $\theta$ are updated through 
% $\theta^{t+1}=\theta^t-\eta\sum_i w_i^{t+1}\nabla\Bar{\mathcal{L}}_i(\theta^t)$. 
\begin{equation}
\small
\label{eqn:tdro_parameter_update}
\theta^{t+1}=\theta^t-\eta\sum_i w_i^{t+1}\nabla\Bar{\mathcal{L}}_i(\theta^t). 
\end{equation}
% \subsubsection{Loss coherence enhancement.}
% \vspace{3pt}
\noindent\textbf{$\bullet$ {Loss consistency enhancement.}}
To alleviate the training instability caused by aggravated data sparsity after group and time period division, we follow~\cite{wen2022distributionally} to keep the streaming estimations of empirical loss: 
% todo appendix 序号
%  of the empirical loss
% The surrogate loss is formulated as
\begin{equation}\notag
\label{eqn:surrogate_loss}
\begin{aligned}
& \Bar{\mathcal{L}}_j^t \leftarrow (1-\mu)\Bar{\mathcal{L}}_j^{t-1} +\mu\Bar{\mathcal{L}}_j^t, \\
\end{aligned}
\end{equation}
% where $\mu$ is the hyper-parameter to control the step size for streaming updates. A smaller $\mu$ leads to more conservative training. 
where $\mu$ is the hyper-parameter to control the streaming step size. A smaller $\mu$ leads to more conservative training (see Appendix~\ref{appendix:streaming} for details).
% **Appendix
% (see Appendix~\ref{appendix:streaming} for details). 

% \begin{center}
% \fcolorbox{black}{gray!6}{\parbox{0.98\linewidth}{\noindent$\bullet$ \textbf{Instantiation.} 
% To instantiate TDRO on cold-start recommender models, we consider applying TDRO based on the overall loss $\mathcal{L}(\theta)$. In detail, we calculate the group weight $\bm{w}$ via Eq. (\ref{eqn:tdro_group_weight_solution}), where $\mathcal{L}(\theta)$ could be 
% substituted by any form of the loss function from cold-start recommender models.
% The training details of TDRO are presented in Algorithm~\ref{algo:TDRO_training}. 
% }}
% \end{center}
% old instantiation
% \vspace{2pt}
% \noindent\textbf{$\bullet$ {Instantiation.}} To instantiate TDRO on cold-start recommender models, we consider applying TDRO based on the overall loss $\mathcal{L}(\theta)$. Specifically, we calculate the group weight $\bm{w}$ via Eq. (\ref{eqn:tdro_group_weight_solution}), where $\mathcal{L}(\theta)$ can be 
% % substituted by Eq. (\ref{eqn:implicit_loss}) and Eq. (\ref{eqn:explicit_loss}) for robust training-based and auxiliary loss-based methods, respectively. 
% substituted by any form of the loss function from cold-start recommender models. 
% Training details of TDRO are presented in Algorithm~\ref{algo:TDRO_training}. 

% \vspace{2pt}
\noindent\textbf{$\bullet$ {Instantiation.}} To instantiate TDRO on cold-start recommender models, we first calculate the group weight $\bm{w}$ via Eq. (\ref{eqn:tdro_group_weight_solution}), where $\mathcal{L}(\theta)$ can be substituted by any form of the loss function from the backend cold-start models. The model parameters will then be optimized based on weighted gradient descent via Eq. (\ref{eqn:tdro_parameter_update}). 
% Specifically, we calculate the group weight $\bm{w}$ via Eq. (\ref{eqn:tdro_group_weight_solution}), where $\mathcal{L}(\theta)$ can be 
% % substituted by Eq. (\ref{eqn:implicit_loss}) and Eq. (\ref{eqn:explicit_loss}) for robust training-based and auxiliary loss-based methods, respectively. 
% substituted by any form of the loss function from cold-start recommender models. 
Training details of TDRO are presented in Algorithm~\ref{algo:TDRO_training}. 

\begin{algorithm}[t]
    \small
    \caption{Training Procedure of TDRO}  
    \label{algo:TDRO_training}
    \begin{algorithmic}[1]

    \Require Number of groups $K$, number of time periods $E$, initial model parameters $\theta^{0}$, initial group weight $\bm{w}=(\frac{1}{K},\frac{1}{K},\dots,\frac{1}{K})$, initial group loss $\Bar{\mathcal{L}}_{i\in[K]}^{0}$, item features $\{\bm{s}_i|i\in\mathcal{I}_w\}$, interactions $\mathcal{D}$, shifting factor strength $\lambda$, period importance $\beta_{e\in[E]}$, weight step size $\eta_w$, streaming step size $\mu$, and learning rate $\eta$.
    % $\alpha$ of $\text{Beta}(\alpha, \alpha)$, and a hyper-parameter $\epsilon$.
    \While{not converge}  
    \ForAll{$i\in\{1,\dots,K\}$}
        % \State Calculate $\Bar{\mathcal{L}}_i^{t+1}(\theta^t;(u,i,y_{ui},\bm{s}_i))$ via the loss function of cold-start recommender model.
         \State Calculate $\Bar{\mathcal{L}}_i^{t}(\theta^t)$ via cold-start loss function.
        \State $\Bar{\mathcal{L}}_i^{t}(\theta^t)\leftarrow(1-\mu)\Bar{\mathcal{L}}_i^{t-1}(\theta^{t-1})+\mu\Bar{\mathcal{L}}_i^{t}(\theta^t)$ 
        % \algorithmiccomment{Loss consistency enhancement}
        \EndFor
    \ForAll{$i\in\{1,\dots,K\}$}
        \State {$w_i^{t+1}\leftarrow w_i^t\exp(\eta_w[(1-\lambda)\Bar{\mathcal{L}}_i^{t}(\theta^{t})+$
        \Statex $\qquad \qquad \qquad \quad$$\lambda(\nabla\Bar{\mathcal{L}}^{t}_i(\theta^{t})\sum\limits_{e=1}^E\sum\limits_{j=1}^K\beta_e\nabla\Bar{\mathcal{L}}_j^{e, t}(\theta^t))])$} 
        % \algorithmiccomment{Group smoothing}
        \EndFor
    \State $w_i^{t+1}\leftarrow w_i^{t+1} / \|\bm{w}^{t+1}\|_1, \forall i\in\{1,\dots,K\}$ \algorithmiccomment{Normalize}
    \State $\theta^{t+1}\leftarrow\theta^{t}-\eta\sum_{i\in[K]}w_i^{t+1}\nabla\Bar{\mathcal{L}}_i^t(\theta^t)$ \algorithmiccomment{Update}
    % \algorithmiccomment{Update parameters}
    
    \EndWhile
    \Ensure Optimized model parameters $\theta$.
	\end{algorithmic}
  
\end{algorithm}

\section{Experiments}
\label{sec:experiment}

We conduct extensive experiments on three real-world datasets to answer the following research questions: 
\begin{itemize}[leftmargin=*]
    \item \textbf{RQ1:} How does our proposed TDRO perform compared to the baselines under temporal feature shifts?
    % \item \textbf{RQ2:} How do the different components of TDRO (\eg two factors for group selection, period importance, and hyper-parameters) affect the performance?
    \item \textbf{RQ2:} How do the different components of TDRO (\ie two factors for group selection) affect the performance?
    \item \textbf{RQ3:} How does TDRO perform over different strengths of temporal feature shifts and how does TDRO mitigate the impact of shifts?
\end{itemize}

\subsection{Experimental Settings}

% \begin{table}[t]
% \setlength{\abovecaptionskip}{0.05cm}
% \setlength{\belowcaptionskip}{0cm}
% \caption{Statistics of three datasets. V and T represent the dimension of visual and textual features, respectively.}
% \label{tab:datasets}
% \setlength{\tabcolsep}{2mm}{
% \resizebox{0.7\textwidth}{!}{
% \begin{tabular}{lccccccc}
% \hline
% \textbf{Dataset} & \textbf{\#User} & \textbf{\#Warm Item} & \textbf{\#Cold Item} & \textbf{\#Interaction} & \multicolumn{1}{c}{\textbf{V}} & \textbf{T} &\textbf{Density} \\ \hline
% \textbf{Amazon} & 21,607 & 75,069 & 18,686 & 169,201 & 64 & - & 0.01\% \\
% \textbf{Micro-video} & 21,608 & 56,712 & 7,725 & 276,629 & 64 & \multicolumn{1}{r}{768} & 0.02\% \\
% \textbf{Kwai} & 7,010 & 74,470 & 12,013 & 298,492 & 64 & - & 0.05\% \\ \hline
% \end{tabular}
% }}
% \vspace{0cm}
% \end{table}

% Please add the following required packages to your document preamble:
% \usepackage{multirow}
\begin{table*}[t]
\renewcommand\arraystretch{1.1}
% \setlength{\abovecaptionskip}{0.05cm}
% \setlength{\belowcaptionskip}{0cm}
% \label{tab:overall}
\setlength{\tabcolsep}{4.5mm}{
\resizebox{\textwidth}{!}{
\begin{tabular}{c|l|lll|lll|lll}
\hline
\multirow{2}{*}{\textbf{Metric}} & \multicolumn{1}{c|}{\multirow{2}{*}{\textbf{Models}}} & \multicolumn{3}{c|}{\textbf{Amazon}} & \multicolumn{3}{c|}{\textbf{Micro-video}} & \multicolumn{3}{c}{\textbf{Kwai}} \\
 & \multicolumn{1}{c|}{} & \multicolumn{1}{c}{\textbf{All}} & \multicolumn{1}{c}{\textbf{Warm}} & \multicolumn{1}{c|}{\textbf{Cold}} & \multicolumn{1}{c}{\textbf{All}} & \multicolumn{1}{c}{\textbf{Warm}} & \multicolumn{1}{c|}{\textbf{Cold}} & \multicolumn{1}{c}{\textbf{All}} & \multicolumn{1}{c}{\textbf{Warm}} & \multicolumn{1}{c}{\textbf{Cold}} \\ \hline\hline
\multirow{14}{*}{\textbf{Recall@20}} & DUIF & 0.0042 & 0.0048 & 0.0129 & 0.0318 & 0.0537 & 0.0771 & 0.0208 & 0.0248 & 0.0158 \\
 & DropoutNet & 0.0050 & 0.0110 & 0.0050 & 0.0187 & 0.0494 & 0.0222 & 0.0099 & 0.0118 & 0.0066 \\
 & M2TRec & 0.0065 & 0.0058 & 0.0068 & 0.0131 & 0.0056 & 0.0298 & 0.0317 & 0.0320 & 0.0009 \\
 & MTPR & 0.0057 & 0.0116 & 0.0082 & 0.0303 & 0.0723 & 0.0542 & 0.0464 & 0.0550 & 0.0049 \\
 & Heater & 0.0065 & 0.0136 & 0.0040 & 0.0469 & 0.1153 & 0.0868 & 0.0452 & 0.0536 & 0.0087 \\
 & CB2CF & 0.0078 & 0.0170 & 0.0074 & 0.0496 & 0.0961 & 0.0928 & 0.0624 & 0.0737 & 0.0064 \\
 & CCFCRec & 0.0071 & 0.0175 & 0.0117 & 0.0435 & 0.0750 & 0.0699 & 0.0098 & 0.0141 & 0.0129 \\
 & InvRL & 0.0120 & 0.0183 & 0.0150 & 0.0578 & 0.0899 & 0.0754 & 0.0588 & 0.0701 & 0.0191 \\ \cline{2-11} 
 & CLCRec & 0.0106 & 0.0200 & 0.0135 & 0.0583 & 0.1135 & 0.0623 & 0.0743 & 0.0884 & 0.0160 \\
 & \textcolor{white}{66}{$+$S-DRO} & 0.0121 & 0.0237 & 0.0144 & 0.0656 & 0.1173 & 0.0719 & 0.0661 & 0.0787 & 0.0172 \\
 % & \textbf{+ TDRO} & \textbf{0.013} & \textbf{0.0237} & \textbf{0.0166} & \textbf{0.0693} & \textbf{0.1165} & \textbf{0.0769} & \textbf{0.0776} & \textbf{0.1023} & \textbf{0.0167} \\ \cline{2-11} 

  & \cellcolor{gray!16}\textcolor{gray!16}{66}\textbf{$+$TDRO} & \cellcolor{gray!16}\textbf{0.0130*} & \cellcolor{gray!16}\textbf{0.0237*} & \cellcolor{gray!16}\textbf{0.0166*} & \cellcolor{gray!16}\textbf{0.0703*} & \cellcolor{gray!16}\textbf{0.1180*} & \cellcolor{gray!16}\textbf{0.0761*} & \cellcolor{gray!16}\textbf{0.0841*} & \cellcolor{gray!16}\textbf{0.1016*} & \cellcolor{gray!16}\textbf{0.0186*} \\ \cline{2-11} 
  
 & GAR & 0.0079 & 0.0200 & 0.0124 & 0.0644 & 0.0962 & 0.0840 & 0.0588 & 0.0706 & 0.0051 \\
 & \textcolor{white}{66}$+$S-DRO & 0.0078 & 0.0189 & 0.0132 & 0.0626 & 0.0894 & 0.0874 & 0.0579 & 0.0690 & 0.0050 \\
 & \cellcolor{gray!16}\textcolor{gray!16}{66}\textbf{$+$TDRO} & \cellcolor{gray!16}\textbf{0.0087*} & \cellcolor{gray!16}\textbf{0.0236*} & \cellcolor{gray!16}\textbf{0.0150*}& \cellcolor{gray!16}\textbf{0.0711*} & \cellcolor{gray!16}\textbf{0.1104*} & \cellcolor{gray!16}\textbf{0.0947*} & \cellcolor{gray!16}\textbf{0.0598*} & \cellcolor{gray!16}\textbf{0.0719*} & \cellcolor{gray!16}\textbf{0.0052} \\ \hline\hline
\multirow{14}{*}{\textbf{NDCG@20}} & DUIF & 0.0020 & 0.0023 & 0.0058 & 0.0204 & 0.0295 & 0.0511 & 0.0158 & 0.0181 & 0.0070 \\
 & DropoutNet & 0.0021 & 0.0043 & 0.0021 & 0.0117 & 0.0286 & 0.0121 & 0.0054 & 0.0061 & 0.0030 \\
 & M2TRec & 0.0032 & 0.0029 & 0.0030 & 0.0075 & 0.0036 & 0.0211 & 0.0247 & 0.0248 & 0.0004 \\
 & MTPR & 0.0029 & 0.0056 & 0.0030 & 0.0175 & 0.0389 & 0.0362 & 0.0324 & 0.0369 & 0.0021 \\
 & Heater & 0.0037 & 0.0075 & 0.0015 & 0.0290 & 0.0653 & 0.0484 & 0.0276 & 0.0312 & 0.0030 \\
 & CB2CF & 0.0037 & 0.0076 & 0.0031 & 0.0254 & 0.0490 & 0.0636 & 0.0446 & 0.0504 & 0.0026 \\
 & CCFCRec & 0.0032 & 0.0074 & 0.0050 & 0.0321 & 0.0410 & 0.0464 & 0.0068 & 0.0092 & 0.0058 \\
 & InvRL & 0.0056 & 0.0079 & 0.0072 & 0.0355 & 0.0493 & 0.0503 & 0.0390 & 0.0444 & 0.0088 \\ \cline{2-11} 
 & CLCRec & 0.0054 & 0.0093 & 0.0061 & 0.0417 & 0.0728 & 0.0444 & 0.0536 & 0.0610 & 0.0071 \\
 & \textcolor{white}{66}{$+$S-DRO} & 0.0060 & 0.0107 & 0.0071 & 0.0451 & 0.0747 & 0.0480 & 0.0472 & 0.0536 & 0.0076 \\
  & \cellcolor{gray!16}\textcolor{gray!16}{66}\textbf{$+$TDRO} & \cellcolor{gray!16}\textbf{0.0066*} & \cellcolor{gray!16}\textbf{0.0112*} & \cellcolor{gray!16}\textbf{0.0077*} & \cellcolor{gray!16}\textbf{0.0507*} & \cellcolor{gray!16}\textbf{0.0794*} & \cellcolor{gray!16}\textbf{0.0511*} & \cellcolor{gray!16}\textbf{0.0597*} & \cellcolor{gray!16}\textbf{0.0719*} & \cellcolor{gray!16}\textbf{0.0081*} \\ \cline{2-11} 
 & GAR & 0.0041 & 0.0088 & 0.0060 & 0.0375 & 0.0496 & 0.0625 & 0.0421 & 0.0485 & 0.0021 \\
 & \textcolor{white}{66}{$+$S-DRO} & 0.0033 & 0.0089 & 0.0052 & 0.0385 & 0.0474 & 0.0532 & 0.0423 & 0.0481 & 0.0021 \\
  & \cellcolor{gray!16}\textcolor{gray!16}{66}\textbf{$+$TDRO} & \cellcolor{gray!16}\textbf{0.0041} & \cellcolor{gray!16}\textbf{0.0110*} & \cellcolor{gray!16}\textbf{0.0066*} & \cellcolor{gray!16}\textbf{0.0419*} & \cellcolor{gray!16}\textbf{0.0571*} & \cellcolor{gray!16}\textbf{0.0638*} & \cellcolor{gray!16}\textbf{0.0431*} & \cellcolor{gray!16}\textbf{0.0495*} & \cellcolor{gray!16}\textbf{0.0024*} \\ \hline
\end{tabular}
}
}
\caption{Overall performance comparison between the baselines and two SOTA models enhanced by TDRO on three datasets. The bold results highlight the better performance in the comparison between the backbone models with and without TDRO. $*$ implies that the improvements over the backbone models are statistically significant ($p$-value \textless 0.01) under one-sample t-tests.}
\label{tab:overall}
\end{table*}

\subsubsection{\textbf{Datasets.}}\label{subsubsec:datasets}
% 我们在三个数据集上进行了实验。
% 分别介绍三个数据集，按照amazon, micro-video， 和kwai的顺序介绍，数据集内容可以参考EQUAL那篇（内容涵盖数据集名字，特点，比如关于什么的推荐，有什么feature等）。
% 讲数据集处理方法，对于前两个数据集：我们 1）sort the interactions according to global timestamps 2）然后按照8:1:1的比例划分到training, validation, 和testing。3）把在training set中的item记做warm item, 在training set中没有出现过的，为cold item。 
% 对于kwai的数据集，由于没有global timestamps, we follow previous work to randomly split the interactions into the training, validation, testing set.

We conducted experiments on three real-world datasets across different domains: 1) \textbf{Amazon}~\cite{he2016ups} is a representative clothing dataset with rich visual features of clothing images. 2) \textbf{Micro-video} is a real-world industry dataset collected from a popular micro-video platform, with rich visual and textual features from thumbnails and textual descriptions. 
% It contains one-month of user interactions with micro-videos, which have rich visual and textual features from thumbnails and textual descriptions. 
3) \textbf{Kwai}\footnote{https://www.kwai.com/.} is a benchmark recommendation dataset provided with rich visual features. 
% For Amazon and Micro-video datasets, we first chronologically sort the interactions according to the timestamps and split them into training, validation, and testing sets with a ratio of 8:1:1. In addition, we divide the items in the validation and testing set into warm and cold sets, where items that do not appear in the training set are regarded as cold items, and the rest as warm items. Regarding the Kwai dataset, due to the lack of global timestamps, we follow previous work~\cite{wei2021contrastive} that randomly split the interactions into the training, validation, and testing sets. The statistics of datasets are summarized in Appendix Table~\ref{tab:datasets}. 
For Amazon and Micro-video datasets, we split the interactions into training, validation, and testing sets chronologically at the ratio of 8:1:1 according to the timestamps. For the Kwai dataset, due to the lack of global timestamps, we instead follow previous work~\cite{wei2021contrastive} that randomly split the interactions. In addition, we divide the items in the validation and testing sets into warm and cold sets, where items that do not appear in the training set are regarded as cold items, and the rest as warm items. 
The statistics of datasets are summarized in Appendix Table~\ref{tab:datasets}.

% \subsubsection{\textbf{Evaluation.}}
% 参考EQUAL：1) full-ranking  2) three different settings: full ranking over all items, warm items only, and cold items only.
% metric使用的是Recall@20 和NDCG@20
% \vspace{2pt}
\noindent\textbf{Evaluation.} 
We adopt the full-ranking protocol~\cite{wei2021contrastive} for evaluation. We consider three different settings: full-ranking over 1) all items, 2) warm items only, and 3) cold items only, denoted respectively as ``all", ``warm", and ``cold'' settings. The widely-used Recall@20 and NDCG@20 are employed as evaluation metrics.

% \vspace{-4pt}
\subsubsection{\textbf{Baselines.}}
% \noindent\textbf{Baselines.} 
% 参考EQUAL，重合baseline可以paraphrase一下EQUAL的
% 我们compare TDRO with representative robust training-based methods ( DUIF, DropoutNet, M2TRec, and MTPR). 以及auxiliary loss-based methods (Heater, CB2CF, CCFCRec, CLCRec, GAR). Besdies，我们比较了TDRO和传统DRO，还考虑了其他potential solution to distribution shift (invariant learning).
% 每个baseline(table2里出现的baseline都列出来) 一一列举，并加引用，用1-2句话解释特点或核心思想（什么做法，达到了什么效果）。
% M2TRec，CCFCRec，和DRO（22www）可以看下他们paper的abstract或者introduction，提炼一下核心思想就可以。
% M2TRec的时候重点介绍一下它是考虑时序信息的，也具备捕捉这种shifting trend的能力；另外DRO也重点强调一下这个是现有利用dro技术用在recommendation的工作
% We conduct a comprehensive comparative analysis of TDRO against state-of-the-art robust training-based methods (DUIF, DropoutNet, M2TRec, and MTPR) and auxiliary loss-based methods (Heater, CB2CF, CCFCRec, CLCRec, GAR). Additionally, we also evaluate TDRO against traditional DRO and other alternative methods (\eg invariant learning) to address the distribution shift problem.
We compare TDRO with competitive cold-start recommender models, including 1) \textit{\textbf{robust training-based methods}}: DUIF~\cite{geng2015learning}, DropoutNet~\cite{volkovs2017dropoutnet}, M2TRec~\cite{shalaby2022m2trec}, and MTPR~\cite{du2020learn}), and 2) \textit{\textbf{auxiliary loss-based methods}}: Heater~\cite{zhu2020recommendation}, CB2CF~\cite{barkan2019cb2cf}, CCFCRec~\cite{zhou2023contrastive}, CLCRec~\cite{wei2021contrastive}, and GAR~\cite{chen2022generative}. Additionally, we also consider 3) \textbf{\textit{potential methods}} to overcome temporal feature shifts: S-DRO~\cite{wen2022distributionally} and invariant learning framework~\cite{du2022invariant}. 
Details of baselines and the hyper-parameter tuning of baselines and TDRO are summarized in Appendix~\ref{appendix:baselines} and~\ref{appendix:hyper-param_setting}.

\begin{table*}[t]
\setlength{\tabcolsep}{2mm}{
\resizebox{\textwidth}{!}{
\begin{tabular}{l|ccc|ccc|ccc}
% \hline
\toprule
 & \multicolumn{3}{c|}{\textbf{Amazon}} & \multicolumn{3}{c|}{\textbf{Micro-video}} & \multicolumn{3}{c}{\textbf{Kwai}} \\ 
 & \textbf{All} & \textbf{Warm} & \textbf{Cold} & \textbf{All} & \textbf{Warm} & \textbf{Cold} & \textbf{All} & \textbf{Warm} & \textbf{Cold} \\ 
\textbf{Methods} & \textbf{Recall@20} & \textbf{Recall@20} & \textbf{Recall@20} & \textbf{Recall@20} & \textbf{Recall@20} & \textbf{Recall@20} & \textbf{Recall@20} & \textbf{Recall@20} & \textbf{Recall@20} \\ \midrule
\textbf{CLCRec} & 0.0106 & 0.0200 & 0.0135 & 0.0583 & 0.1135 & 0.0623 & 0.0743 & 0.0884 & 0.0160 \\
\textbf{w/o Worst-case Factor} & 0.0121 & 0.0219 & 0.0157 & 0.0648 & 0.1138 & 0.0687 & 0.0790 & 0.0997 & 0.0145 \\
\textbf{w/o Shifting Factor} & 0.0126 & 0.0228 & 0.0160 & 0.0643 & 0.1145 & 0.0622 & 0.0797 & 0.0986 & 0.0165 \\
\textbf{TDRO} & \textbf{0.0130} & \textbf{0.0237} & \textbf{0.0166} & \textbf{0.0703} & \textbf{0.1180} & \textbf{0.0761} & \textbf{0.0814} & \textbf{0.1016} & \textbf{0.0186} \\ \bottomrule
\end{tabular}
}}
\caption{Ablation study of worst-case factor and shifting factor. The best results are highlighted in bold.}
\label{tab:ablation}
\end{table*}
\subsection{Overall Performance (RQ1)}

% Overall Performance
The overall performance of the baselines and the two SOTA cold-start methods equipped with S-DRO and TDRO is reported in Table~\ref{tab:overall}, from which we can observe the following:
\noindent$\bullet$ 
    % Auxiliary loss-based methods (Heater, CB2CF, CCFCRec, CLCRec, GAR) typically outperform the robust training-based ones (DUIF, DropoutNet, M2TRec, MTPR).  
    % The reason is that robust training-based methods directly utilize feature representations to fit interactions, which inevitably introduces noises and hurt the learning of the recommender model. 
    Auxiliary loss-based methods typically outperform the robust training-based ones.  
    The reason is that robust training-based methods directly utilize feature representations to fit interactions, which inevitably introduces noises and hurt the learning of the recommender model. 
    % Meanwhile, auxiliary loss-based methods decouple the CF and feature representations space, which leads to 1) protecting CF representations from feature noise for robust interaction prediction on warm items; and 2) improving cold performance effectively by designing different auxiliary loss (\eg contrastive loss) for alignment. 
    Meanwhile, auxiliary loss-based methods decouple the CF and feature representations space, which protects the CF representations from feature noises and improves cold performance effectively via different auxiliary losses. 

    % \vspace{1pt}
    % \noindent$\bullet$ CLCRec consistently yields impressive performance across three datasets. The superior results are attributed to the integration of contrastive loss for aligning feature and CF representations, where mutual information between feature and CF space is maximized to learn a more robust feature extractor. Besides, GAR exhibits competitive performance despite its instability. The superior performance of GAR indicates the effectiveness of adversarial constraint for similar distributions of CF and feature representations. 
    \noindent$\bullet$ CLCRec consistently yields impressive performance across the three datasets. This is attributed to the integration of contrastive loss for aligning feature and CF representations, where mutual information between feature and CF space is maximized for robust prediction. Besides, by introducing adversarial constraints for similar distributions of CF and feature representations, GAR exhibits competitive performance despite its instability. 

    \noindent$\bullet$ In most cases, S-DRO improves the performance of cold items compared to the backbone model. The stable improvements are attributed to the tail performance guarantee over potential shifted distributions, which may partially cover the shifted cold item distribution. In addition, our proposed TDRO consistently outperforms S-DRO and the backbone model on all and cold performance by a large margin. 
    This justifies the effectiveness of TDRO in enhancing the generalization ability of the feature extractor. Moreover, capturing the shifting patterns is also helpful for achieving steady improvements for warm items, reflecting the superiority of TDRO in alleviating the temporal feature shifts issue. 
    Possible reasons for inferior performance on InvRL and M2TRec are discussed in Appendix~\ref{appendix:overall_performance}.

\subsection{In-depth Analysis}
% In this subsection, we study each component of TDRO, \ie the worst-case and shifting factor. In addition, we investigate the effectiveness of TDRO under different strengths of item feature shifts and analyze how EQUAL mitigates the impact of item feature shifts\footnote{We present the experimental results of TDRO instantiated on CLCRec and omit the results with similar observations of TDRO on GAR to save space.}. 
% In this subsection\footnote{We focus on the analysis of TDRO instantiated on CLCRec due to its better performance compared to GAR to save space. And the hyper-parameter analysis can be found in Appendix~\ref{appendix:hyper-analysis}.}, we study two factors of TDRO, investigate the effectiveness of TDRO under different strengths of temporal feature shifts, and explore how EQUAL mitigates the impact of temporal feature shifts. 
In this subsection, 
% \footnote{We focus on the analysis of TDRO instantiated on CLCRec due to its better performance compared to GAR to save space.}, 
we study two factors of TDRO, investigate the effectiveness of TDRO under different strengths of temporal feature shifts, and explore how EQUAL mitigates the impact of temporal feature shifts. 
% In this subsection, we study two factors of TDRO, investigate the effectiveness of TDRO under different strengths of temporal feature shifts, and explore how EQUAL mitigates the impact of temporal feature shifts. 

\subsubsection{\textbf{Ablation Study (RQ2).}}
To study the effectiveness of the worst-case and shifting factor, we implement TDRO without (w/o) each factor, separately. From Table~\ref{tab:ablation}, we can find that: 
1) The performance declines if either the worst-case factor or the shifting factor is removed. This verifies the effectiveness of incorporating the optimization over worst-case group and the performance improvements for all groups based on the shifting trend.  
% 2) Meanwhile, removing each factor still outperforms CLCRec under ``all" setting, showing the superiority of each factor for enhancing generalization ability. 
2) Removing each factor still outperforms CLCRec (``all" setting). 
% This is because the two factors alleviate the item feature shifts from two aspects, \ie reducing the impact of shifts for uniformly good performance and capturing the shifts for better prediction, respectively. 
 % This is because worst-case factor encourages robustness under distribution shifts while shifting factor utilizes the shifting trend to alleviate item feature shifts. 
This indicates that either performance lower bound guarantee or leveraging shifting trends improves generalization ability. 
% 3) The inferior cold performance of removing the worst-case factor compared to CLCRec is probably due to the different dataset pre-processing for Kwai, which results in less significant shifting trends (\cf Section~\ref{subsubsec:datasets}).

% 

% \vspace{-2pt}
% \subsubsection{\textbf{Effect of period importance (RQ2). }}
% % todo: 图要不要加一个dro的基准线 - 不加了
% To analyze the impact of period importance $\beta_e=\exp(p \cdot e)$ on capturing shifting trend, we vary the steepness control factor $p$ from 0.05 to 1 and report the results on Amazon in Figure~\ref{fig:hp_effect}(a). The results with similar observations on Micro-video and Kwai are omitted to save space. From the figures, we can find that 
% 1) the performance increases as we enlarge $p$. This is because a smaller $p$ intends to pay attention to each time period uniformly whereas a larger $p$ encourages to emphasize the time periods closer to the test phase, pushing the model to capture the shifting trend of item features of subsequent cold items. 
% 2) Due to that only the last time period is considered in TDRO as $p$ approaches infinity, blindly increasing $p$ overlooks other relatively earlier time periods containing useful information to capture shifting patterns, thus hurting the learning of recommender models.
% \vspace{3pt}

\begin{table}[t]
\setlength{\tabcolsep}{1.8mm}{
\resizebox{0.48\textwidth}{!}{
\begin{tabular}{l|ccc|ccc}
\toprule
% \hline
 & \multicolumn{3}{c|}{\textbf{Amazon}} & \multicolumn{3}{c}{\textbf{Micro-video}} \\
& \textbf{Group1} & \textbf{Group2} & \textbf{Group3} & \textbf{Group1} & \textbf{Group2} & \textbf{Group3} \\ \midrule
\textbf{Distance} & 48 & 62 & 123 & 13 & 19 & 39 \\ \midrule
\textbf{CLCRec} & 0.0218 & 0.0075 & 0.0024 & 0.1131 & 0.0503 & 0.0116 \\ 
\textbf{TDRO} & \textbf{0.0254} & \textbf{0.0110} & \textbf{0.0027} & \textbf{0.1321} & \textbf{0.0598} & \textbf{0.0139} \\ \bottomrule
\end{tabular}
}}
\caption{Recall@20 over user groups with different strengths of temporal feature shifts under ``all'' setting. }
\label{tab:user_group}

\end{table}

\subsubsection{\textbf{User Group Evaluation (RQ3).}}
% todo: 画一个柱状图，下面标一下每组的average distance，加一下DRO的，说一下similar observation on kwai, 或者加一组kwai上的, - ok
We further inspect how TDRO performs under different strengths of temporal feature shifts by evaluating TDRO on different user groups. 
Specifically, we calculate the Euclidean distance of the average item features between the training set and testing set for each user. Next, we rank the users according to the distance, and then split the users into three groups (denoted as Group 1, Group 2, and Group 3) based on the ranking. The results \wrt Recall@20 is given in Table~\ref{tab:user_group}. Despite that the performance of both CLCRec and TDRO declines gradually as the shifts become more significant, TDRO consistently outperforms CLCRec in each group, validating the effectiveness of TDRO in enhancing the generalization ability to temporal feature shifts. 

% 1) from Group 1 to Group 3, the performance of CLCRec and TDRO gradually declines. This is reasonable because it is hard to predict interactions accurately under more significant item feature shifts. 
% 2) Nevertheless, TDRO consistently outperforms CLCRec in each group, validating the effectiveness of TDRO in alleviating the item feature shifts. 

\begin{table}[t]
\setlength{\tabcolsep}{3.5mm}{
\resizebox{0.48\textwidth}{!}{
\begin{tabular}{l|cc|cc}
\toprule
 & \multicolumn{2}{c|}{\textbf{All}} & \multicolumn{2}{c}{\textbf{Cold}} \\
 & \textbf{Worst-case} & \textbf{Popular} & \textbf{Worst-case} & \textbf{Popular} \\ \midrule
\textbf{CLCRec} & 0.0166 & 0.0168 & 0.0088 & 0.0088 \\
\textbf{TDRO} & \textbf{0.0173} & \textbf{0.0195} & \textbf{0.0123} & \textbf{0.0125} \\ \bottomrule
\end{tabular}
}}
\caption{Recall@20 of the item group with the worst performance and the item group of top 25\% popular items.}
\label{tab:item_group}
\end{table}
\subsubsection{\textbf{Item Group Analysis (RQ3).}} 
To explore how TDRO alleviates the impact of temporal feature shifts, we analyze the generalization ability enhancement of TDRO on Amazon \wrt item groups. 
In detail, we calculate the item popularity (\ie interaction proportion) in the testing set and divide the items into four subgroups based on the popularity scores. We then conduct evaluation on each item subgroup to see whether TDRO: 1) guarantees the worst-case group performance, and 2) enhances the performance over the group with the top 25\% popular items. 
% We conduct item group evaluation under both ``all'' and ``cold'' settings with the results reported in Table~\ref{tab:item_group}. 
As shown in Table~\ref{tab:item_group}, the boosted performance on worst-case group and popular items partially explains the superior performance of TDRO. 

\section{Conclusion and Future Work}
\label{sec:conclusion}
In this work, we revealed the critical issue of temporal item feature shifts in the cold-start recommendation.  
% , the key lies in strengthening generalization ability of the feature extractor over temporally shifted distributions. 
To overcome this issue, we proposed a novel temporal DRO learning framework called TDRO, which 
% 1) considering worst-case performance for tail performance guarantee, 
% 1) considers the worst-case performance to confer the guarantee of the performance lower bound, 
1) considers the worst-case performance for the performance lower bound guarantee, 
and 2) leverages the shifting trend of item features to enhance the performance of popular groups in subsequent cold items. 
% We conducted extensive experiments on three real-world datasets under various settings to validate the effectiveness of TDRO in achieving robust prediction under temporal item feature shifts. 
Empirical results on three real-world datasets validated the effectiveness of TDRO in achieving robust prediction under temporal item feature shifts. 
% 怎么做的，达到了什么效果/为了达到什么效果，我们做了什么

% This work uncovers the underlying problem of inappropriate cold-start item recommendation: temporal feature shifts, leaving many promising directions to be explored in the future. 
This work highlights the temporal item feature shifts in cold-start recommendation and extends DRO to alleviate the shifts, leaving many promising directions to be explored in the future. 
% 1) adaptive environment importance function, for more accurate capturing of the shifting trend
One is to consider adaptive environment importance for more fine-grained modeling of the shifting trend. 
% 2) 
Moreover, it is worthwhile to explore more effective group division strategies beyond the pre-defined ones, to fulfill the potential of TDRO in enhancing the model's generalization ability enhancement. 
Another promising direction is to leverage large language models to capture nuanced item semantics for improving cold-start recommendation~\cite{bao2023bi,bao2023large,wang2023generative}. 
% instantiate TDRO for improving generalization ability in other tasks, such as
\clearpage
{
\fontsize{9.0pt}{10.0pt} \selectfont
\bibliography{aaai24}
}
\clearpage
\appendix
\section{Appendix}
\label{sec:appendix}
\subsection{DRO}\label{appendix:DRO}
% \noindent\textbf{$\bullet $ Discrepancy-based DRO.} 
% DRO is an effective solution that could achieve consistently high performance across various distribution shifts~\cite{zhou2021examining,duchi2018learning}. Technically, it optimizes the worst-case performance over the pre-defined uncertainty set (\ie potential shifted distributions) for controlling the performance lower bound. In formal, 
\subsubsection{Overpessimism issue.}
DRO is an effective solution that could achieve consistently high performance across various distribution shifts~\cite{zhou2021examining,duchi2018learning}. In formal, 
\begin{equation}
\small
\label{appeqn:dro_optimization}
\begin{aligned}
% & \theta^*_{\text{DRO}} := \mathop{\arg\min}_{\theta\in\Theta} \left\{\mathcal{R}(\theta):=\mathop{\sup}_{Q\in\mathcal{Q}}\mathbb{E}_{(u,i,y_{ui})\in{Q}} [\mathcal{L}\left(\theta;(u,i, y_{ui},\bm{s}_i)\right)]\right\}, \\
& \theta^*_{\text{DRO}} := \mathop{\arg\min}_{\theta\in\Theta} \left\{\mathop{\sup}_{Q\in\mathcal{Q}}\mathbb{E}_{(u,i,y_{ui})\in{Q}} [\mathcal{L}\left(\theta;(u,i, y_{ui},\bm{s}_i)\right)]\right\}, \\
\end{aligned}
\end{equation}
% where $\mathcal{R}(\theta)$ is the worst-case risk
where  
the uncertainty set $\mathcal{Q}$ encodes the possible shifted distributions that we want our model to perform well on. 

The uncertainty set $\mathcal{Q}$ is typically defined as a ball surrounding the training distributions endowed with a certain discrepancy metric, such as f-divergence~\cite{duchi2018learning} and Wasserstein Distance~\cite{liu2022distributionally}. 
% By carefully choosing the discrepancy metric and the radius for determining the boundary of the possible shifted distributions, we could confer the robustness of the model to a wide range of distribution shifts. 
By choosing the discrepancy metric and the radius for determining the boundary of the possible shifted distributions, we could confer the robustness of the model to a wide range of distribution shifts. 
Nevertheless, such a definition of uncertainty set could lead to the over-pessimism issue, where implausible shifted distributions are being overwhelmingly considered~\cite{liu2022distributionally,oren2019distributionally}. Therefore, to overcome the over-pessimism issue, Group-DRO is proposed to find the potential shifted distributions on a group level~\cite{oren2019distributionally,sagawa2020distributionally,hu2018does}.

\subsubsection{Objective of Group-DRO.} 
In Group-DRO, the training distribution $\mathcal{P}$ is assumed to be a mixture of $K$ pre-defined groups 
$\{P_i|i=1,\dots,K\}$, \ie $\mathcal{P}:=\sum_{i=1}^{K}w_iP_i$, where $w_i$ is the mixture ratio of the group $i$. 
Then, the uncertainty set of Group-DRO is defined as $\mathcal{Q}:=\{\sum_{i=1}^{K}w'_iP_i: w'_i\in\Delta_K\}$, 
where $\Delta_K$ is the $K-1$-dimensional probability simplex. 
Since the maximum of $\sum_{i=1}^{K}w'_iP_i$ is attained at a vertex due to linearity, the objective of Group-DRO is reformulated as 
% the worst-case risk $\mathcal{R}(\theta)$ of Group-DRO can be reformulated as 
% \begin{equation}
% \label{eqn:group_worstcase_risk}
% \begin{aligned}
% % & {\hat{\mathcal{R}}(\theta)}:=\mathop{\max}_{g\in\mathcal{G}}\mathbb{E}_{(u,i,y_{ui})\sim P_g}[\mathcal{L}\left(\theta;(u,i, y_{ui},\bm{s}_i)\right)]. \\
% & {\hat{\mathcal{R}}(\theta)}:=\mathop{\max}_{j\in\{1,\dots,K\}}\mathbb{E}_{(u,i,y_{ui})\sim P_j}[\mathcal{L}\left(\theta;(u,i, y_{ui},\bm{s}_i)\right)]. \\
% \end{aligned}
% \end{equation}
% To alleviate the item feature shifts issue, DRO is a potential solution
% DRO 
% DRO 和 Group DRO
% Then, by substituting Eq. (\ref{eqn:group_worstcase_risk}) into Eq. (\ref{eqn:dro_optimization}), we obtain the objective of Group-DRO, 
\begin{equation}
\small
\label{appeqn:gdro_optimization}
\begin{aligned}
% & \theta^*_{\text{Group-DRO}} := \mathop{\arg\min}_{\theta\in\Theta} \
% \left\{\hat{\mathcal{R}} (\theta):=\mathop{\max}_{j\in\{1,\dots,K\}}\mathbb{E}_{(u,i,y_{ui})\sim P_j} [\mathcal{L}\left(\theta;(u,i, y_{ui},\bm{s}_i)\right)]\right\}. \\
\theta^*_{\text{DRO}} := \mathop{\arg\min}_{\theta\in\Theta} \left\{
\mathop{\max}_{j\in[K]}\mathbb{E}_{(u,i,y_{ui})\sim P_j} [\mathcal{L}\left(\theta;(u,i, y_{ui},\bm{s}_i)\right)]
\right\}. 
\end{aligned}
\end{equation}

\subsection{Formula Derivation}\label{appendix:derivation}

% \noindent\textbf{$\bullet$ Interpretation of shifting factor.} 
\subsubsection{Interpretation of shifting factor.} 

Via First-order Taylor, we can rewrite Eq. (\ref{eqn:tdro_group_selection_1}) into,
\begin{equation}
\small
\label{eqn:derive_taylor}
% \left\{
\begin{aligned}
j^*&\approx\mathop{\arg\min}_{j\in\{1,\dots,K\}} {(\lambda-1)\Bar{\mathcal{L}}_j(\theta^t)} + \\
&\qquad{\lambda\sum\limits_{e=1}^E\sum\limits_{i=1}^K\beta_e [\mathcal{L}_i^e(\theta^t)+\eta\nabla_\theta {\mathcal{L}}^e_i(\theta^t)(\theta^{t+1}-\theta^t)]} \\
&= \mathop{\arg\min}_{j\in\{1,\dots,K\}} (\lambda-1) \Bar{\mathcal{L}}_j(\theta^t) + \lambda\sum\limits_{e=1}^E\sum\limits_{i=1}^K\beta_e [\mathcal{L}_i^e(\theta^t) - \eta \langle \bm{g}^e_i,{\bm{g}}_j\rangle] ,\\
&= \mathop{\arg\min}_{j\in\{1,\dots,K\}} (\lambda-1) \Bar{\mathcal{L}}_j(\theta^t) - \lambda\sum\limits_{e=1}^E\sum\limits_{i=1}^K\Tilde{\beta}_e \langle \bm{g}^e_i,{\bm{g}}_j\rangle, \\
% &= \mathop{\arg\max}_{j} (1-\lambda) \Bar{\mathcal{L}}_j(\theta^t) + \lambda\Bar{\bm{g}}_j^T\sum\nolimits_e\sum\nolimits_i\Tilde{\beta}_e\bm{g}_i^e, \\
\end{aligned}
% \right.
\end{equation}
% \begin{multline}
% \label{eqn:derive_taylor}
% % \left\{
% \begin{aligned}
% j^*&\approx\mathop{\arg\min}_{j\in\{1,\dots,K\}} {(\lambda-1)\Bar{\mathcal{L}}_j(\theta^t)} + \\{\lambda\sum\nolimits_{e=1}^E\sum\nolimits_{i=1}^K\beta_e [\mathcal{L}_i^e(\theta^t)+\eta\nabla_\theta {\mathcal{L}}^e_i(\theta^t)(\theta^{t+1}-\theta^t)]} \\
% &= \mathop{\arg\min}_{j\in\{1,\dots,K\}} (\lambda-1) \Bar{\mathcal{L}}_j(\theta^t) + \\
% \lambda\sum\nolimits_{e=1}^E\sum\nolimits_{i=1}^K\beta_e [\mathcal{L}_i^e(\theta^t) - \eta \langle \bm{g}^e_i,{\bm{g}}_j\rangle] ,\\
% &= \mathop{\arg\min}_{j\in\{1,\dots,K\}} (\lambda-1) \Bar{\mathcal{L}}_j(\theta^t) \\
% - \lambda\sum\nolimits_{e=1}^E\sum\nolimits_{i=1}^K\Tilde{\beta}_e \langle \bm{g}^e_i,{\bm{g}}_j\rangle, \\
% % &= \mathop{\arg\max}_{j} (1-\lambda) \Bar{\mathcal{L}}_j(\theta^t) + \lambda\Bar{\bm{g}}_j^T\sum\nolimits_e\sum\nolimits_i\Tilde{\beta}_e\bm{g}_i^e, \\
% \end{aligned}
% % \right.
% \end{multline}
where $\Tilde{\beta}_e=\beta_e \cdot \eta$. We use $\beta_e$ to denote $\Tilde{\beta}_e$ in Eq. (\ref{eqn:tdro_group_selection_taylor}) for notation brevity. 

% \noindent\textbf{$\bullet$ Group weight smoothing.} 
\subsubsection{Group weight smoothing.} 
We consider the optimization problem, 
\begin{equation}
\small
\label{eqn:derive_kkt}
\begin{aligned}
&\underset{w}{\min}-\sum_i w_ic_i+\frac{1}{\eta_w}\sum_iw_i \log\frac{w_i}{w_i^t}+\delta_{\ge0}(\bm{w}), \\
&\mathrm{s.t.} \quad \sum_iw_i=1. \\
\end{aligned}
\end{equation}
where $c_i=(1-\lambda)\Bar{\mathcal{L}_i}(\theta^t) +\lambda\langle \bm{g}_i,\sum_{e=1}^E\sum_{j=1}^K\beta_e \bm{g}^e_j\rangle$, and $\delta_{\ge0}(\bm{w})$ is the indicator function of set $\{x|x\ge0\}$. For $y\in\mathbb{R}$, we have the Lagrangian,
\begin{equation}
\small
\begin{aligned}
L(\bm{w},y)&=\sum_i\left[ w_ic_i+ \frac{1}{\eta_w}\sum_iw_i \log\frac{w_i}{w_i^t}+\delta_{\ge0}(\bm{w}) + w_iy\right] -y. \\
\end{aligned}
\end{equation}
Then the Lagrange dual function is
$\theta_y=\max L(\bm{w},y)$. For every $i$, we solve,
\begin{equation}
\small
\begin{aligned}
&\max -w_ic_i + \frac{1}{\eta_w} \log\frac{w_i}{w_i^t} + w_iy, \\
&\mathrm{s.t.} \quad\sum_iw_i=1.\\
\end{aligned}
\end{equation}
By setting $\nabla_{w_i}=-c_i+ \frac{1}{\eta_w}(\log\frac{w_i}{w_i^t} + 1) +y=0 $, we obtain the solution as,
\begin{equation}
\begin{aligned}
% w_i=w_i^t\exp(\eta_wc_i)\exp(\eta_wy). \\
w_i=\frac{w_i^t\exp(\eta_wc_i)}{\sum_jw_j^t\exp(\eta_wc_j)}.\\
\end{aligned}
\end{equation}

\subsubsection{Streaming estimations of empirical loss.}\label{appendix:streaming}
% todo
% \todo{TODO}
In DRO training, the loss variance for the selected group can be high~\cite{wen2022distributionally}, thus leading to instability. To reduce the loss variance, previous work~\cite{wen2022distributionally} proposes a streaming algorithm. The key idea is to keep streaming estimations of the empirical loss $\Bar{\mathcal{L}}_j^t$ by updating the loss value in a small step $\mu$, in a similar way to stochastic gradient descent: 
\begin{equation}\notag
\label{app_eqn:surrogate_loss}
\begin{aligned}
& \Bar{\mathcal{L}}_j^t \leftarrow (1-\mu)\Bar{\mathcal{L}}_j^{t-1} +\mu\Bar{\mathcal{L}}_j^t. \\
\end{aligned}
\end{equation}
We can find that a smaller $\mu$ preserves more loss value at the last time step, \ie $\Bar{\mathcal{L}}_j^{t-1}$. As such, the optimization will be less affected by batches where sparse subgroups do not exist.

\subsection{Micro-video Dataset}\label{appendix:dataset}
Micro-video is an industrial dataset collected from a worldwide micro-video APP, which contains user-item interactions on extensive micro-videos from October 15 to November 14, 2021. 
This dataset consists of rich side information, especially the content feature of items. The content features of items (\ie micro-videos) include visual features, such as thumbnails of different resolutions. Besides, each micro-video has diverse textual features, such as titles, descriptions, and tags in multiple languages. 

Though Amazon and Kwai are two popular benchmark datasets for recommendation in different domains, they only have visual features for extracting feature representations of cold items. Therefore, 
we conduct experiments on Micro-video datasets, to validate the effectiveness of TDRO on different modalities of item features. 

\subsection{Baselines}\label{appendix:baselines}
% We compare TDRO with competitive cold-start recommender models, including the robust training-based methods (DUIF, DropoutNet, M2TRec, and MTPR), and the auxiliary loss-based methods (Heater, CB2CF, CCFCRec, CLCRec, and GAR). Additionally, we also consider methods that have the potential to overcome the issue of distribution shifts (DRO and invariant learning framework). 
\begin{itemize}
    \item \textbf{DUIF}~\cite{geng2015learning} discards CF representations and solely utilizes feature representations to align with interactions. 
    \item \textbf{DropoutNet}~\cite{volkovs2017dropoutnet} leverages feature and CF representations for training interaction predictor, where item CF representations are randomly removed to strengthen the alignment between feature representations and interactions. 
    \item \textbf{M2TRec}~\cite{shalaby2022m2trec} is a sequential-based model which integrates a Transformer encoder to learn feature representations, enabling the ability to capture the shifting trend. Besides, it removes CF representations in training and inference.
    \item \textbf{MTPR}~\cite{du2020learn} replaces some warm item CF representations with zero vectors for robust learning in a multi-task manner. 
    \item \textbf{Heater}~\cite{zhu2020recommendation} utilizes a mixture of experts to extract personalized feature representations, and achieves alignment with interactions by minimizing the distance between feature representations and pre-trained CF representations.
    \item \textbf{CB2CF}~\cite{barkan2019cb2cf} introduces a general feature extractor to align feature representations with CF representations learned from interactions via MSE loss.
    \item \textbf{CCFCRec}~\cite{zhou2023contrastive} encourages the feature extractor to learn robust feature representations by minimizing the distance between feature representations of co-occurring warm items. 
    \item \textbf{CLCRec}~\cite{wei2021contrastive} aligns feature and CF representations effectively by maximizing the mutual information between feature and CF representations via contrastive loss. 
    \item \textbf{GAR}~\cite{chen2022generative} uses adversarial training to bridge the distribution gap between feature and CF representations. 
    \item \textbf{InvRL}~\cite{du2022invariant} leverages the invariant learning framework, which pays attention to learning invariant feature representations that can achieve robust interaction prediction across different distributions. 
    \item \textbf{S-DRO}~\cite{wen2022distributionally} is a method that employs DRO to mitigate the performance gap between user groups in recommender systems. We implement S-DRO on both CLCRec and GAR.
\end{itemize}
% dataset
\subsection{Hyper-parameters Settings}\label{appendix:hyper-param_setting}
% \subsection{Statistics of datasets.}
% \section{Hyper-parameter Settings}
% \subsubsection{\textbf{Hyper-parameter Settings.}}
% 第一段参考EQUAL，paraphrase一下就行，内容是一致的。
% 每个baseline的调参范围参考EQUAL
% M2TRec, CCFCRec, DRO，和TDRO的调参范围参考excel
% xinyu todo: 整理新baseline调参范围-ok
% To ensure a fair comparison, we set the same dimension of CF and feature representations to 128 for all methods. Moreover, we adopt a two-layer MLP with a hidden size of 256 as the feature extractor to generate feature representations. To obtain the optimal performance, we perform a hyper-parameter tuning process for the learning rate and weight decay, where we search in the range of \{1e−4,1e−3,1e−2\} and \{1e−4, 1e−3, 1e−2, 1e−1\}, respectively. We select the default hyper-parameters based on the Recall metric evaluated on the validation set under the ``all'' setting. Moreover, we conduct a more focused search for other specific hyper-parameters, and their search scopes are as follows:
To ensure a fair comparison, we set the same dimension of CF and feature representations to 128 and adopt a two-layer MLP with a hidden size of 256 as the feature extractor for all methods. Besides, we conduct hyper-parameters tuning, where the best hyper-parameters are selected based on the Recall over the validation set under ``all'' setting. We tune the learning rate and weight decay in the range of $\{1e^{-4},1e^{-3},1e^{-2}\}$ and $\{1e^{-4}, 1e^{-3}, 1e^{-2}, 1e^{-1}\}$, respectively, for all baselines. And we search other model-specific hyper-parameters with searching scopes and the best ones as follows: 

\begin{figure*}[t]
  \centering 
  \hspace{-0.2in}
  \subfigure{
    \includegraphics[height=1.45in]{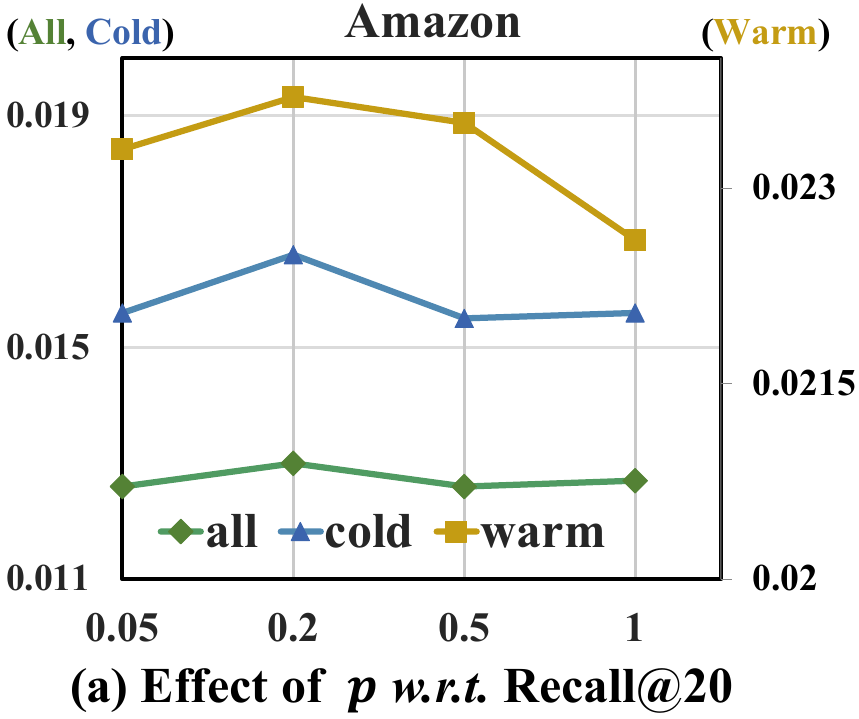}} 
  \hspace{-0.105in}
  \subfigure{
    \includegraphics[height=1.45in]{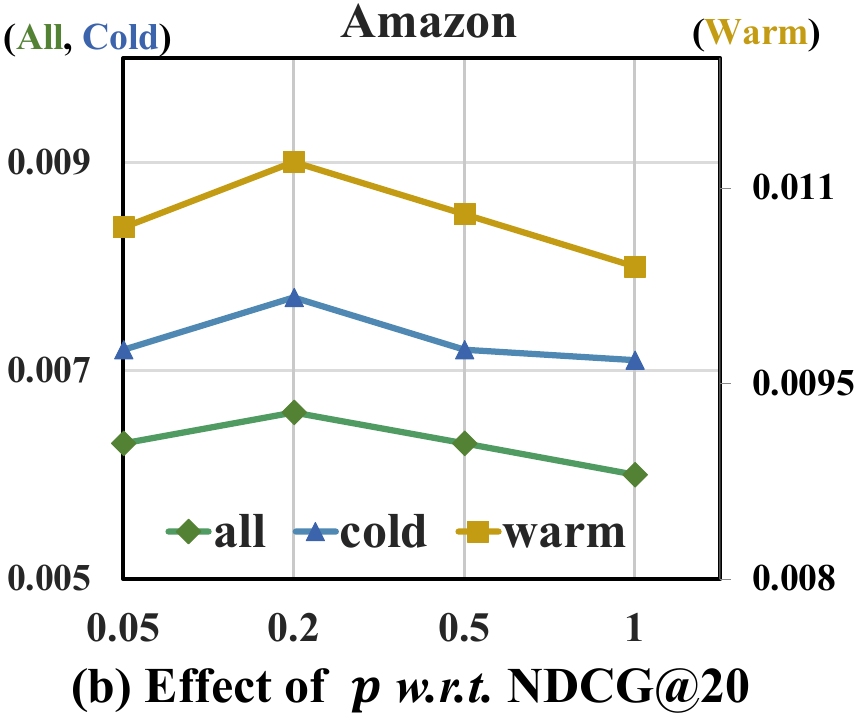}} 
  \hspace{-0.105in}
  \subfigure{
    \includegraphics[height=1.45in]{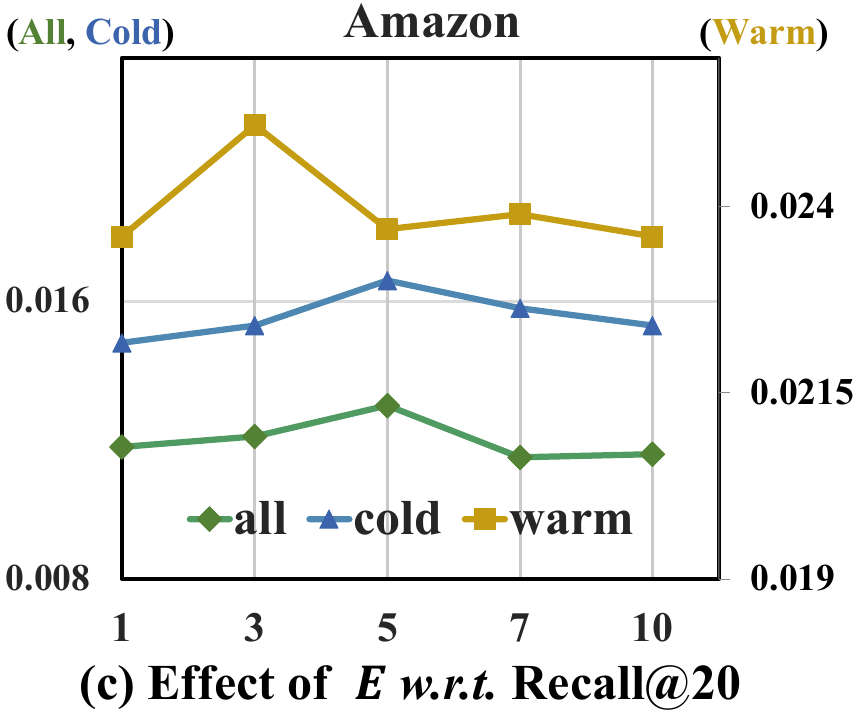}} 
  \hspace{-0.105in}
  \subfigure{
    \includegraphics[height=1.45in]{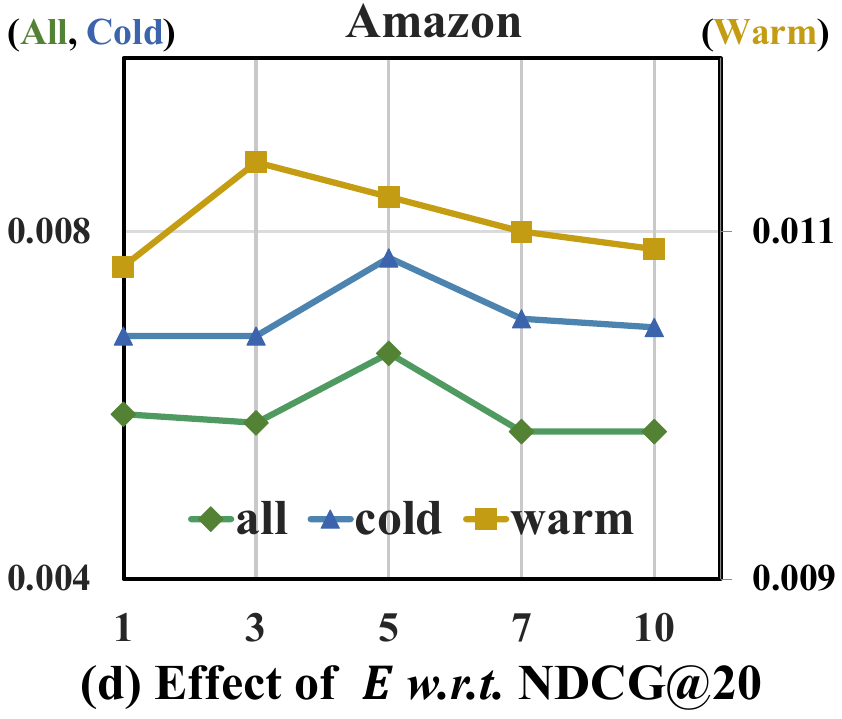}} 
  % \caption{(a) and (b) shows performance over user groups with different strengths of item feature shifts. Results are presented \wrt Recall@20 under ``all" setting. Performance of worst-case item group and the item group of top 25\% popular items are given in (c) and (d).}
    \caption{Effect of period importance steepness control factor $p$ for period importance and period number $E$.} 
  \label{fig:appendix_hp_p_E}
  % \vspace{-0.30cm}
\end{figure*}

\begin{itemize}[leftmargin=*]
    \item \textbf{DUIF.} The best learning rate is $1e^{-3}$ on the three datasets. 
    The best weight decay is $1e^{-3}$ on Micro-video and Kwai, and $1e^{-4}$ on Amazon. 
    \item \textbf{DroupNet}. The dropout ratio is tuned in $\{0.2,0.5,0.8\}$. 
    The best dropout ratio is $0.8$ on Amazon, and $0.5$ on Micro-video and Kwai. 
    \item \textbf{M2TRec}. We tune the number of Transformer layers and attention heads in $\{1,2,4\}$ and $\{1,2,4,8\}$, separately. 
    Both of the best numbers of Transformer layers and attention heads are $2$ on the three datasets. 
    \item \textbf{CCFCRec, CLCRec, CB2CF, and Heater}. The coefficient of the auxiliary loss is searched in $\{0.1,0.3,0.5,0.7,0.9\}$. As for CCFCRec, the numbers of positive and negative samples for auxiliary loss are tuned in $\{1,2,4\}$ and $\{10,20,40\}$, respectively. Besides, we choose the ratio for CLCRec to randomly replaces CF representations with feature representations in $\{0.1,0.5,0.9\}$. 
    As for CCFCRec, the best coefficients of the auxiliary loss are $0.5$, $0.3$, and $0.1$ on Amazon, Micro-video, and Kwai, respectively. The best numbers of positive and negative samples are $1$ and $10$ on the three datasets. 
    As for CLCRec, the best coefficient of auxiliary loss is $0.1$ on the three datasets, and the ratio for random replacing is $0.5$, $0.9$, and $0.3$ on Amazon, Micro-video, and Kwai, respectively. 
    \item \textbf{GAR}. We separately search the coefficients of adversarial loss and interaction prediction loss in $\{0.1,0.5,0.9\}$. 
    The best coefficients of adversarial loss are $0.5$, $0.9$, and $0.9$ on Amazon, Micro-video, and Kwai, respectively. The best coefficients of interaction prediction are $0.6$, $0.9$, and $0.9$ on Amazon, Micro-video, and Kwai, respectively.
    \item \textbf{InvRL}. The number of environments, the coefficient of environment constraint, and the strength of mask regularization are chosen from $\{5,10,20,30\}$, $\{0.1,0.5,1\}$, and $\{0.1,0.5,1\}$, respectively. 
    The best number of environments, the coefficient of environment constraint, and the strength of mask regularization are $10$, $1$, and, $0.1$, respectively, on the three datasets. 
    
    \item \textbf{S-DRO}. We select the group number $K$ from $\{1,3,5,7,10\}$. The streaming step size $\mu$ and the regularization strength for group importance smoothing $\eta_w$ are searched in $\{0.1,0.2,0.3,0.5\}$ and $\{0.01,0.1,0.2,0.3,0.5\}$, respectively. 
    As for the CLCRec backbone model, the best group numbers are $3$, $3$, and $5$ on Amazon, Micro-video, and Kwai, respectively. The best streaming step size and the regularization strength are $0.2$ and $0.1$, respectively on the three datasets. 
    As for the GAR backbone model, the best group numbers are $5$, $3$, and $5$ on Amazon, Micro-video, and Kwai, respectively. The best streaming step size and the regularization strength are $0.2$ and $0.1$, respectively on the three datasets. 
    \item \textbf{TDRO}. We choose the time period number $E$, the shifting factor strength $\lambda$, and the steepness control factor $p$ for period importance are searched in $\{1,3,5,7,10\}$, $\{0.1,0.3,0.5,0.7\}$ and $\{0.05,0.2,0.5,1\}$, respectively. The searching scopes for other shared hyper-parameters are consistent with S-DRO. 
    As for the CLCRec backbone model, the best $K$ is $5$, $3$, and $3$ on Amazon, Micro-video, and Kwai, respectively. The best $E$ is $5$, $3$, and $5$ on Amazon, Micro-video, and Kwai, respectively. The best $\lambda$ is $0.3$, $0.1$, and $0.3$ on Amazon, Micro-video, and Kwai, respectively. The best $p$ is $0.2$ on the three datasets. 
    As for the GAR backbone model, the best $K$ is $3$ on the three datasets. The best $E$ is $3$, $3$, and $5$ on Amazon, Micro-video, and Kwai, respectively. The best $\lambda$ is $0.9$, $0.1$, and $0.3$ on Amazon, Micro-video, and Kwai, respectively. The best $p$ is $0.2$ on the three datasets.  
\end{itemize}

The statistics of three datasets for experiments are summarized in Table~\ref{tab:datasets}.
\begin{table}[t]
\small
\begin{center}
\setlength{\tabcolsep}{1.6mm}{
\resizebox{0.48\textwidth}{!}{
\begin{tabular}{lccccccc}
\toprule
\textbf{Dataset} & \textbf{\#User} & \textbf{\#Warm} & \textbf{\#Cold} & \textbf{\#Int} & \multicolumn{1}{c}{\textbf{V}} & \textbf{T} &\textbf{Density} \\ \midrule
\textbf{Amazon} & 21,607 & 75,069 & 18,686 & 169,201 & 64 & - & 0.01\% \\
\textbf{Micro-video} & 21,608 & 56,712 & 7,725 & 276,629 & 64 & \multicolumn{1}{r}{768} & 0.02\% \\
\textbf{Kwai} & 7,010 & 74,470 & 12,013 & 298,492 & 64 & - & 0.05\% \\ \bottomrule
\end{tabular}
}}
\end{center}
\caption{Statistics of three datasets. ``Int'' denotes ``Interactions'', ``V'' and ``T'' represent the dimension of visual and textual features, respectively.}
\label{tab:datasets}
\vspace{0cm}
\end{table}

\subsection{Additional Analysis of Experimental Results}\label{appendix:overall_performance}
% \noindent$\bullet\quad$\textbf{Overall performance.}   
\subsubsection{Overall performance.} 
From Table~\ref{tab:overall}, we may find that as for potential solutions to item feature shifts, InvRL surpasses its backbone model (CLCRec) on cold performance whereas it sacrifices the accuracy of warm item recommendation. This is reasonable since InvRL captures the invariant part of CF and feature representations, which encourages robust cold item recommendation. Nevertheless, it overlooks the variant part, which may be essential for interaction prediction for warm items. 
In contrast, M2TRec, which leverages the sequential shifting pattern, usually has inferior performance compared to other baselines. The reason might be that it discards both user and item CF representations for prediction, thus being severely affected by the negative impact of feature noise for capturing sequential patterns.  

% \noindent$\bullet\quad$\textbf{Ablation study performance.} 
\subsubsection{Ablation study.} 
In Table~\ref{tab:ablation}, the inferior cold performance of removing the worst-case factor compared to CLCRec is probably due to the different dataset pre-processing for Kwai, where the interactions are randomly split into the training, validation, and testing sets. As such, the shifting trends of Kwai might be less significant (\cf Section~\ref{subsubsec:datasets}).

\subsection{Hyper-parameter Analysis}\label{appendix:hyper-analysis}
\subsubsection{\textbf{Effect of period importance. }}
% todo: 图要不要加一个dro的基准线 - 不加了
To analyze the impact of period importance $\beta_e=\exp(p \cdot e)$ on capturing shifting trend, we vary the steepness control factor $p$ from 0.05 to 1 and report the results on Amazon in Figure~\ref{fig:appendix_hp_p_E}(a) and (b). The results with similar observations on Micro-video and Kwai are omitted to save space. From the figures, we can find that 
1) the performance increases as we enlarge $p$. This is because a smaller $p$ intends to pay attention to each time period uniformly whereas a larger $p$ encourages to emphasize the later time periods closer to the test phase, pushing the model to capture the shifting trend of item features of subsequent cold items. 
2) Due to that only the last time period is considered in TDRO as $p$ approaches infinity, blindly increasing $p$ overlooks other relatively earlier time periods containing useful information to capture shifting patterns, thus hurting the learning of cold-start recommender models.

\subsubsection{\textbf{Effect of time period number. }}
We vary the time period number from 1 to 10 and present the results in Figure~\ref{fig:appendix_hp_p_E}(c) and (d), where we can find that: 1) Slightly increasing time period numbers yields better performance. This is because a larger time period number leads to a more fine-grained time period division, facilitating TDRO to capture the shifting trend of item features accurately. Nevertheless, 2) the time period number cannot be too large as it will exacerbate the interaction data sparsity for each time period and cause training instability, thus degrading the performance. 

\subsubsection{\textbf{Effect of shifting trend strength.}}
\begin{figure}[t]
  \centering 
  \hspace{-0.2in}
  \subfigure{
    \includegraphics[height=1.4in]{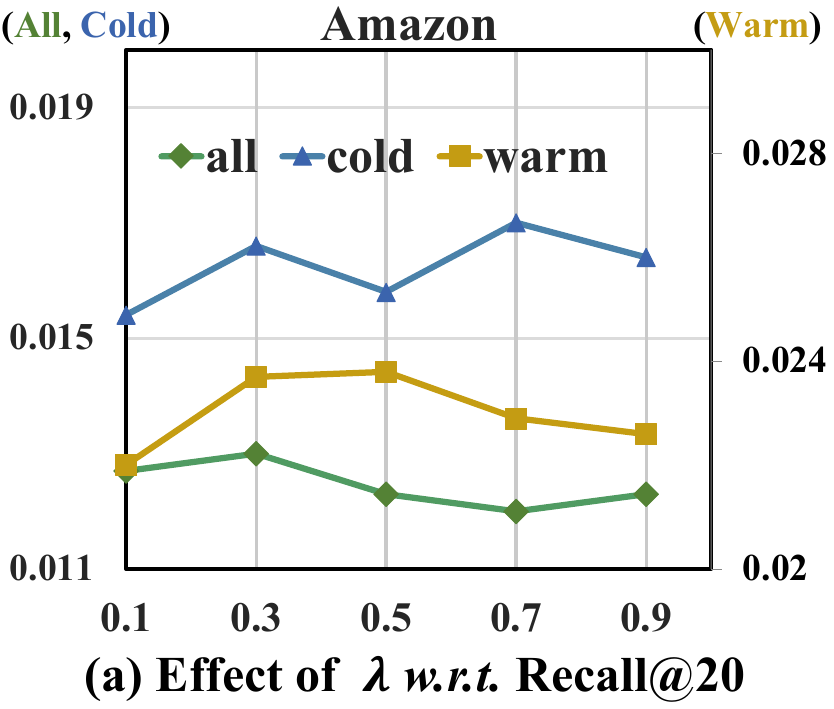}} 
  % \hspace{-0.125in}
    \hspace{-0.08in}
  % \subfigure{
  %   \includegraphics[height=1.6in]{figures/hp_env_amazon_R.pdf}}
  %   \hspace{-0.105in}
  \subfigure{
    \includegraphics[height=1.4in]{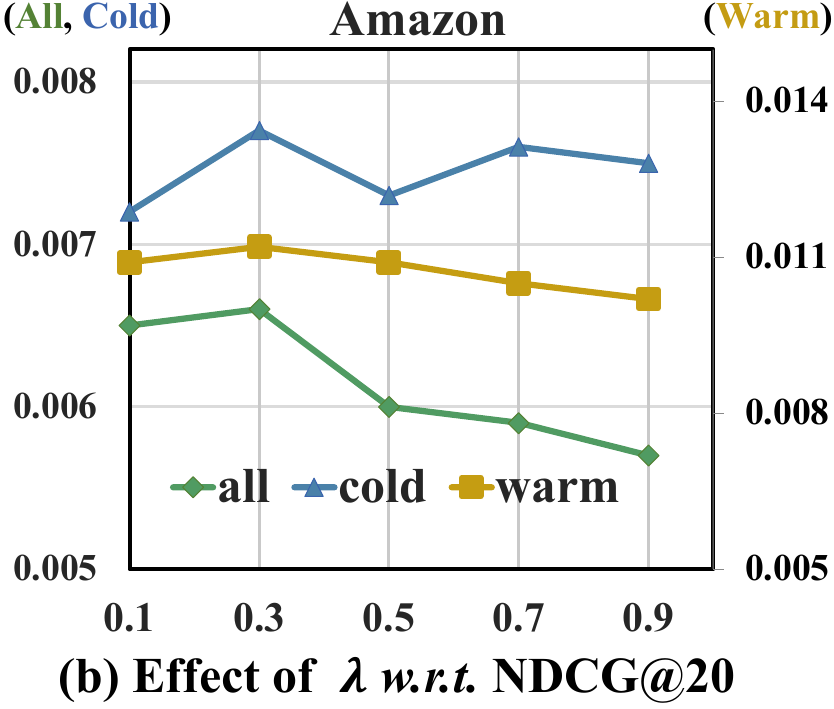}}
    \caption{Effect of the strength of shifting factor $\lambda$.}
  \label{fig:hp_effect}
\end{figure}

We inspect the effect of the shifting factor by changing $\lambda$ from 0.1 to 0.9. As shown in Figure~\ref{fig:hp_effect}, it is observed that stronger incorporation of the shifting trend typically intends to yield better performance on cold items, indicating the importance of considering shifting patterns in robustness enhancement over cold items. However, the all and warm performance declines if we consider the shifting factor too much (\ie worst-case factor is weak), which is probably due to the overlook of the minority group of warm items. Therefore, we should carefully choose $\lambda$ to balance the trade-off between the worst-case group and the groups that are likely to become popular in cold items. 
% todo time period的recall，ndcg图
\begin{figure}[t]
% \vspace{-0.2cm}
% \setlength{\abovecaptionskip}{-0.1cm}
% \setlength{\belowcaptionskip}{-0.3cm}
  \centering 
  \hspace{-0.2in}
  \subfigure{
    \includegraphics[height=1.4in]{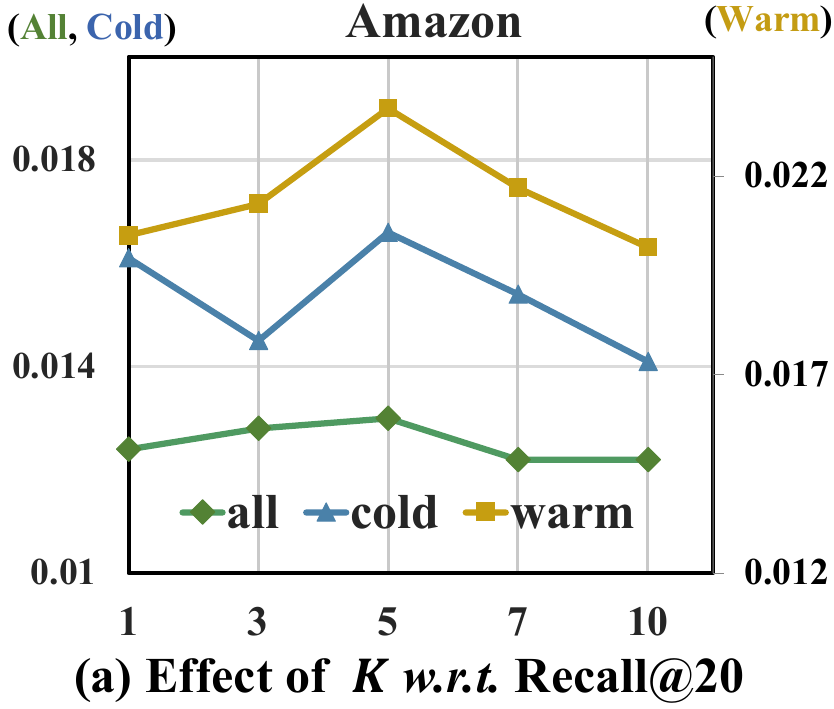}} 
  \hspace{-0.105in}
  \subfigure{
    \includegraphics[height=1.4in]{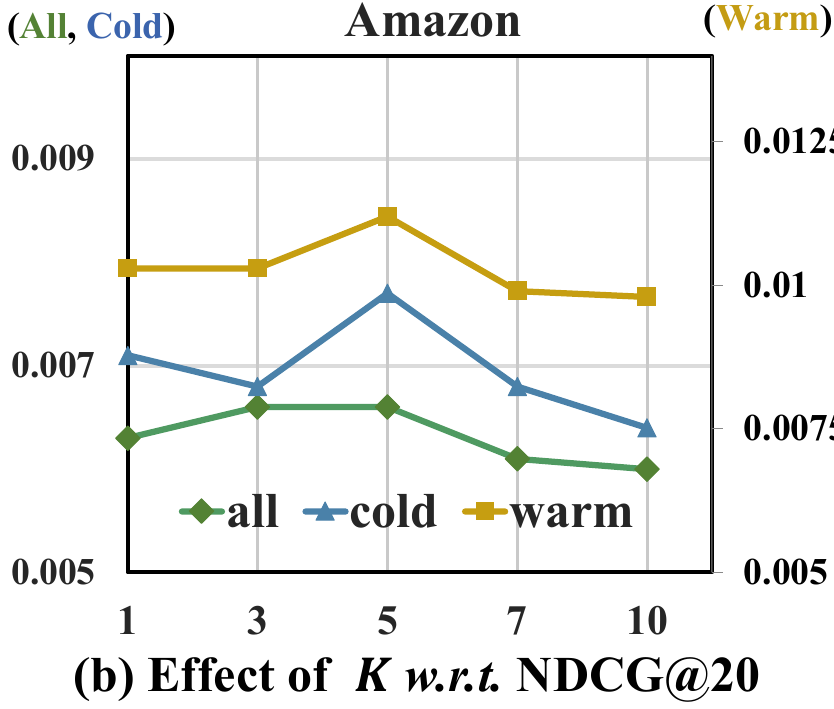}} 
    \caption{Effect of group number $K$.} 
  \label{fig:appendix_hp_K}
  % \vspace{-0.30cm}
\end{figure}

\subsubsection{\textbf{Effect of group number. }}
We change $K$ from 1 to 10, with the results reported in Figure~\ref{fig:appendix_hp_K}. We can observe that the performance achieves the best when $K=5$ under ``all'', ``warm'', and ``cold'' settings. The fluctuating performance from $K=1$ to $K=10$ indicates that the performance is not proportional to the group number. This makes sense since it is non-trivial to recover the group-level preference aligning well with real-world scenarios (\eg a variety of clothing). A more efficient way of choosing an appropriate group number from interactions can be explored in future works. 

\subsection{Detailed Related Work}\label{appendix:related_work}
\subsubsection{Cold-start Recommendation.}
Traditional CF methods typically rely on historical user-item interactions to learn CF representations for interaction prediction~\cite{wang2021denoising}. However, the influx of cold items obstructs traditional CF methods from providing appropriate recommendations due to the lack of historical interactions~\cite{zhao2022improving,rajapakse2022fast,raziperchikolaei2021shared,pulis2021siamese,du2022socially}, escalating imbalanced impressions on the item side~\cite{huan2022industrial,zhu2021fairness,sun2021form,wang2021privileged,chu2023meta}. 
% As for cold users, their interests can be effectively explored by recommending popular items to them. 
% Randomly recommending numerous cold items to learn item CF representations is inefficient and may hurt the user experience. 
% Therefore, it is crucial to solve the cold-start item issue and achieve accurate interaction prediction over cold items. 
Hence, it is crucial to enhance the cold-start item recommendation~\cite{houlsby2014cold,pan2019warm,vartak2017meta,neupane2022dynamic,wei2020fast}. 

% Existing methods typically address the cold-start problem by aligning the extracted feature representations with CF counterparts in the following ways: (1) Alignment via robust training, where alignment is achieved by randomly dropping out the CF representation of warm items for robust prediction purely based on feature representation. For example, DropoutNet~\cite{volkovs2017dropoutnet} employs a technique of randomly dropping CF representations of warm items for more robust training, which can help to fill the gap and compensate for the missing CF information for cold items.  Instead of dropping CF representations, MTPR~\cite{du2020learn} uses a method of replacing warm item CF representations with zero vectors and applies multi-task learning for more robust training. (2) Alignment via auxiliary loss which explicitly constrain the distance between CF representation and feature representation by the auxiliary loss. For example, CLCRec~\cite{wei2021contrastive} maps the item feature into the feature representation and pushes it to be closer to the CF space via contrastive learning. GAR~\cite{chen2022generative} utilizes adversarial training to learn a robust feature extractor, bridging the gap between CF space and feature representation space.
Existing methods address the cold-start problem by aligning the extracted feature representations with user-item interactions~\cite{meng2020wasserstein,shi2018attention,guo2017integration,hao2021pre,togashi2021alleviating}, typically falling into two research lines: 
1) Robust training-based methods, where both feature and CF representations are utilized to predict interactions while CF representations are randomly corrupted to encourage robust alignment between feature representations and interactions. 
% For example, DropoutNet~\cite{volkovs2017dropoutnet} randomly drops CF representations of warm items for robust training, which encourages robust interacation predictions over cold items solely based on feature representations. MTPR~\cite{du2020learn} instead replaces CF representations of warm items with zero vectors and applies multi-task learning for robust training.
For example, ~\cite{volkovs2017dropoutnet} randomly drops CF representations of warm items for robust training. 
% and~\cite{du2020learn} replaces CF representations of warm items with zero vectors and applies multi-task learning for robust training. 
% Instead of dropping CF representations, MTPR~\cite{du2020learn} uses a method of replacing warm item CF representations with zero vectors and applies multi-task learning for more robust training. 
2) Auxiliary loss-based methods achieve alignment by introducing different auxiliary loss for minimizing the distance between feature representations and CF representations learned from interactions. 
For example,~\cite{wei2021contrastive} maximizes the mutual information between feature and CF representations spaces via contrastive loss, and~\cite{chen2022generative} utilizes adversarial training to effectively bridge the gap between feature and CF spaces.

However, previous methods suffer from temporal feature shifts from warm items to cold items, where the feature representations of cold items may not be well captured by the feature extractor learned from warm items. 
In this work, we highlight the importance of strengthening the generalization ability of the feature extractor against temporal feature shifts, where the key lies in considering temporally shifted distributions that reflect the shifting trend of item features to achieve robust interaction prediction. 

\subsubsection{Distributionally Robust Optimization.}
DRO is an effective approach that aims to achieve uniform performance against distribution shifts by optimizing the worst-case performance over a pre-defined uncertainty set~\cite{rahimian2019distributionally,michel2022distributionally}. 
The most representative line of research works lies in discrepancy-based DRO that defines the uncertainty set as a ball surrounding the training distributions with different discrepancy metrics (\eg f-divergence~\cite{duchi2018learning}, MMD ball~\cite{staib2019distributionally}, and Wasserstein Distance~\cite{liu2022distributionally}). 
For example,~\cite{zhai2021doro} leverages the Cressie-Read family of R$\Acute{\text{e}}$nyi divergence to define the uncertainty set, and~\cite{liu2022distributionally} is developed based on Wasserstein Distance.
Due to the fact that discrepancy-based DRO unnecessarily considers implausible distributions (\ie over-pessimism issue~\cite{oren2019distributionally,sagawa2020distributionally,duchi2023distributionally}), 
% suffers from the over-pessimism issue, where implausible distributions are unnecessarily considered, 
another line falls into Group-DRO~\cite{zhou2021examining,goel2020model}. Works along this line define the uncertainty set as a set of mixtures of subgroups in the training set, which encourages DRO to focus on meaningful distribution shifts. 
% On the other hand, Group DRO is the new emerging method that defines the uncertainty set as the combination of subgroups of the training set, aiming to optimize the group-level worst-case performance during the training process. 
% For instance, ~\cite{wen2022distributionally}  proposed a streaming algorithm based on Group DRO to reduce the uncertainty of loss estimation. 
% In addition, ~\cite{piratla2022focus} selects the subgroup that leads to the largest reduction in global loss after gradient descent compared with Group DRO rather than the minority group. 
% For instance, PDRO~\cite{michel2021modeling} models the uncertainty set using generative models and uses Importance Sampling to ensure that the data is sampled from real data.  However,  this method introduces an extra number of parameters and has more computational costs.
For instance,~\cite{oren2019distributionally} focuses on the distribution shifts between topics for natural language modeling, and~\cite{wen2022distributionally} optimizes the worst-case performance over different user groups in recommender systems.
% Recently, there are also some parameter-based methods that define the uncertainty set as parametric generative models. 
Recently, some works have been proposed to define the uncertainty set based on generative models~\cite{michel2021modeling}, which inevitably introduces extra parameters, resulting in a burden of high computational costs. 

However, applying DRO directly to cold start recommendations has inconsistency issues. The overemphasis on the minority group of warm items may weaken the expressiveness of the groups that are likely to become popular in the upcoming cold items, thus limiting the robustness enhancement for cold item recommendation. In this work, we propose to leverage the temporal shifting trend to guide DRO to improve the generalization ability of the feature extractor against temporal feature shifts. 
% Besides, potential solution for  compared and discussed in Section~\ref{sec:experiment}.  

Other literature on robustness has also been extensively studied. 
Invariant learning~\cite{arjovsky2019invariant,du2022invariant,koyama2020out,liu2021heterogeneous} considers the invariant part robust to distribution shifts but it overlooks the variant part, which may be essential for prediction. 
Re-weighting loss~\cite{zhang2021causal,kim2021premere,yang2023dgrec} assigns weights to samples, which however relies heavily on correlations between group density and task difficulty and yields inferior performance than DRO (\cf Table 1 in ~\cite{wen2022distributionally}).

\end{document}